\documentclass[pra,twocolumn,showpacs,superscriptaddress]{revtex4-1}

\usepackage[utf8]{inputenc}
\usepackage[english]{babel}
\usepackage[T1]{fontenc}

\usepackage{graphicx}
\usepackage{amsmath}
\usepackage{amsfonts}
\usepackage{amssymb}

\usepackage{amsthm}
\usepackage{bbm}

\usepackage{makecell}

\usepackage[breaklinks=true,colorlinks]{hyperref}

\newcommand{\cf}{cf.\ }

\newcommand{\coloneq}{\mathrel{\mathop:}=}
\newcommand{\eqcolon}{=\mathrel{\mathop:}}
\newcommand{\dd}{\mathrm{d}}
\newcommand{\Tr}{\operatorname{Tr}}
\newcommand{\realt}{\operatorname{Re}}
\newcommand{\bkew}[3]{\left\langle{#1}\middle|{#2}\middle|{#3}\right\rangle}
\newcommand{\ket}[1]{\left|{#1}\right\rangle}
\newcommand{\ketbra}[2]{\left|{#1}\middle\rangle\middle\langle{#2}\right|}
\newcommand{\proj}[1]{\ketbra{#1}{#1}}
\newcommand{\ew}[1]{\left\langle{#1}\right\rangle}

\newcommand{\kB}{k_\mathrm{B}}
\newcommand{\sminus}{\sigma_-}
\newcommand{\splus}{\sigma_+}
\newcommand{\sminusbar}{\bar{\sigma}_-}
\newcommand{\splusbar}{\bar{\sigma}_+}
\newcommand{\sminustilde}{\tilde{\sigma}_-}
\newcommand{\splustilde}{\tilde{\sigma}_+}
\newcommand{\indexc}{\mathrm{c}}
\newcommand{\indexh}{\mathrm{h}}
\newcommand{\Gc}{G_\indexc}
\newcommand{\Gh}{G_\indexh}
\newcommand{\betaeff}{\beta_\mathrm{eff}}
\newcommand{\betac}{\beta_\indexc}
\newcommand{\betah}{\beta_\indexh}
\newcommand{\Teff}{T_\mathrm{eff}}
\newcommand{\Tc}{T_\indexc}
\newcommand{\Th}{T_\indexh}
\newcommand{\Jc}{J_\indexc}
\newcommand{\Jh}{J_\indexh}
\newcommand{\psib}{\psi_\mathrm{b}}
\newcommand{\psid}{\psi_\mathrm{d}}
\newcommand{\Neff}{N_\mathrm{eff}}
\newcommand{\Pid}{\Pi_\mathrm{d}}
\newcommand{\Pidew}{\ew{\Pi_\mathrm{d}}_{\rho(0)}}
\newcommand{\Omegacrit}{\Omega_\mathrm{crit}}

\begin{document}

\title{Performance limits of multilevel and multipartite quantum heat machines}

\author{Wolfgang Niedenzu}
\affiliation{Department of Chemical Physics, Weizmann Institute of Science, Rehovot~7610001, Israel}

\author{David Gelbwaser-Klimovsky}
\affiliation{Department of Chemical Physics, Weizmann Institute of Science, Rehovot~7610001, Israel}
\affiliation{Department of Chemistry and Chemical Biology, Harvard University, Cambridge, MA~02138, USA}

\author{Gershon Kurizki}
\affiliation{Department of Chemical Physics, Weizmann Institute of Science, Rehovot~7610001, Israel}

\begin{abstract}
We present the general theory of a quantum heat machine based on an $N$-level system (working medium) whose $N-1$ excited levels are degenerate, a prerequisite for steady-state interlevel coherence. Our goal is to find out: To what extent is coherence in the working medium an asset for heat machines? The performance bounds of such a machine are common to (reciprocating) cycles that consist of consecutive strokes and continuous cycles wherein the periodically driven system is constantly coupled to cold and hot heat baths. Intriguingly, we find that the machine's performance strongly depends on the relative orientations of the transition-dipole vectors in the system. Perfectly aligned (parallel) transition dipoles allow for steady-state coherence effects, but also give rise to dark states, which hinder steady-state thermalization and thus reduce the machine's performance. Similar thermodynamic properties hold for $N$ two-level atoms conforming to the Dicke model. We conclude that level degeneracy, but not necessarily coherence, is a thermodynamic resource, equally enhancing the heat currents and the power output of the heat machine. By contrast, the efficiency remains unaltered by this degeneracy and adheres to the Carnot bound.
\end{abstract}

\date{August 12, 2015}

\maketitle

\section{Introduction}

The extensive efforts over the years to reconcile thermodynamics with quantum mechanics~\cite{gemmerbook,mahlerbook,alicki1979quantum,kosloff1984quantum,kosloff2013quantum} have not yet fully resolved the fundamental question: What is truly quantum about quantum thermodynamics? Is there more to it than just restating traditional thermodynamic principles for quantized systems? Attempts to cope with this problem have revolved around possible quantum resources that may boost the thermodynamic performance of heat machines as compared to their classical counterparts. A prime contender for such a resource is quantum coherence~\cite{agarwal2001quantum,scully2003extracting,kozlov2006inducing,deffner2013information,dorfman2013photosynthetic,tscherbul2014long,uzdin2015quantum}. Intriguing schemes have predicted power and/or efficiency increase in quantum heat engines~\cite{scully2003extracting} due to sustainable (steady-state) coherence in non-thermal baths, as well as coherence-enhanced performance of photovoltaic solar heat converters~\cite{scully2010quantum,scully2011quantum,svidzinsky2011enhancing,svidzinsky2012enhancing,creatore2013efficient} or coherent effects in photosynthesis~\cite{dorfman2013photosynthetic,dijkstra2015coherent}. According to an interesting view~\cite{deffner2013information}, coherence in thermodynamics acts as an information reservoir or Maxwell's demon, i.e., as an extra resource that can tip the entropy balance and the division of input energy between heat and work in favor of the latter~\cite{szilard1929ueber,landauer1961irreversibility}.

\par

Here we explore the thermodynamics of multilevel systems with excited-state degeneracy, which is a prerequisite for the persistence of interlevel quantum coherence in steady state imposed by \emph{thermal baths}~\cite{agarwal2001quantum,kozlov2006inducing,tscherbul2014long,gelbwaser2014power}. Our goal is to address the question: To what extent is coherence an asset when such systems are employed as working media in heat machines? 

\par

We show that the bounds on the efficiency and power output in the presence of coherence are general and common to all cycles, i.e., reciprocating cycles that consist of consecutive strokes and continuous cycles. Similar thermodynamic performance bounds are shared by multilevel systems whose excited states are degenerate and transition dipoles are perfectly aligned, and by multipartite Dicke systems~\cite{dicke1954coherence}. These general considerations (Sec.~\ref{sec_H_SB}) are followed (Secs.~\ref{sec_system}--\ref{sec_ss}) by a study of a quantum heat machine containing a working medium that has $N-1$ degenerate upper levels, whose transition dipoles to the ground state may not all be aligned. The system is constantly coupled to two spectrally distinct, hot and cold, thermal baths and is periodically modulated (Stark-shifted) by an external field. This modulation acts as a piston in the heat machine~\cite{gelbwaser2013minimal}. Our objective is to study the steady-state operation of such a heat machine, i.e., the limit cycle of its dissipative evolution, by deriving its heat currents, power, and efficiency in the ususal regime of weak system-bath coupling~\cite{gemmerbook,mahlerbook,alicki1979quantum,kosloff1984quantum,kosloff2013quantum,carmichaelbook,breuerbook,gorini1976completely,lindblad1976generators} (Sec.~\ref{sec_heat_currents}).

\par 

The present theory extends our previous study of heat machines whose working medium is a periodically modulated two-level system (TLS) that is continuously coupled to two (hot and cold) baths~\cite{gelbwaser2013minimal,alicki2014quantum,gelbwaser2014power} (see also Ref.~\cite{gelbwaser2015thermodynamics} for a recent review). The merit of such models is that they are amenable to a full quantum-mechanical analysis by the Floquet method~\cite{alicki2012periodically,kolar2012quantum,gelbwaser2013minimal}. Moreover, their continuous-cycle operation avoids the difficulty of ensuring compatibility with the laws of thermodynamics, in contrast to machines that are operated via reciprocating cycles that consist of four strokes (e.g., the Carnot- or the Otto cycle), where, alternately, only a hot or a cold bath is coupled to the working medium at any time~\cite{schwablbook}. The difficulty therein is to properly account for the highly nonadiabatic energy and entropy flows induced by frequent on- and off-switching of interactions with alternate heat baths in consecutive strokes. By contrast, continuous-cycle, periodically-driven, heat machines can in a straightforward manner be made compatible with the first and second laws of thermodynamics regardless of their nonadiabaticity~\cite{gelbwaser2013minimal,kosloff2013quantum}.

\par

Steady-state coherences between upper levels are shown here to arise as a consequence of the thermalization of the initial state under the condition of strict transition-dipole alignment. However, thermalization may then be partially blocked, i.e., be incomplete, by the mere presence of dark states. For this reason we introduce the notion of \emph{thermalization capability} of the initial state. We show that this capability should be maximized, by avoiding the initial population of dark states, in order to cause maximal power enhancement by the multilevel heat machine, as compared to its two-level counterpart. We argue that it is the thermalization capability rather than steady-state coherence that underlies the power-enhancement mechanism: The key resource of the multilevel heat machine is the thermalization of transitions that share a \emph{common} ground state, and are thereby correlated, whether coherently or incoherently. Similar principles govern heat machines based on a multiatom Dicke system, notwithstanding their different dynamics (Sec.~\ref{sec_dicke}).

\par

The heat currents and the power are analytically shown (Secs.~\ref{sec_heat_currents} and~\ref{sec_dicke}) to be both strongly boosted by the same enhancement factor with respect to a single two-level system. Consequently, the efficiency of the multilevel or multipartite heat machine equals that of its two-level counterpart and thus adheres to the Carnot efficiency bound, which is reached at zero power---the operating point at which the quantum heat engine is transformed into a quantum refrigerator~\cite{gelbwaser2013minimal}. The Curzon--Ahlborn limit~\cite{curzon1975efficiency} for the efficiency of classical heat engines at maximum power, however, can be surpassed for appropriately engineered bath spectra and carefully chosen temperatures~\cite{gelbwaser2013minimal}.

\par

In Sec.~\ref{sec_realizations} we discuss possible realizations of the pertinent models and their limitations. Notably, we point out that the $N$-atom Dicke system may be the most straightforward experimental implementation of the effects predicted in this work. In Sec.~\ref{sec_conclusions_and_outlook} the conclusions of this work are presented.

\section{Thermodynamic r\^ole of coherences and dark states}\label{sec_H_SB}

\subsection{Model}

\par
\begin{figure}
  \centering
  \includegraphics[width=\columnwidth]{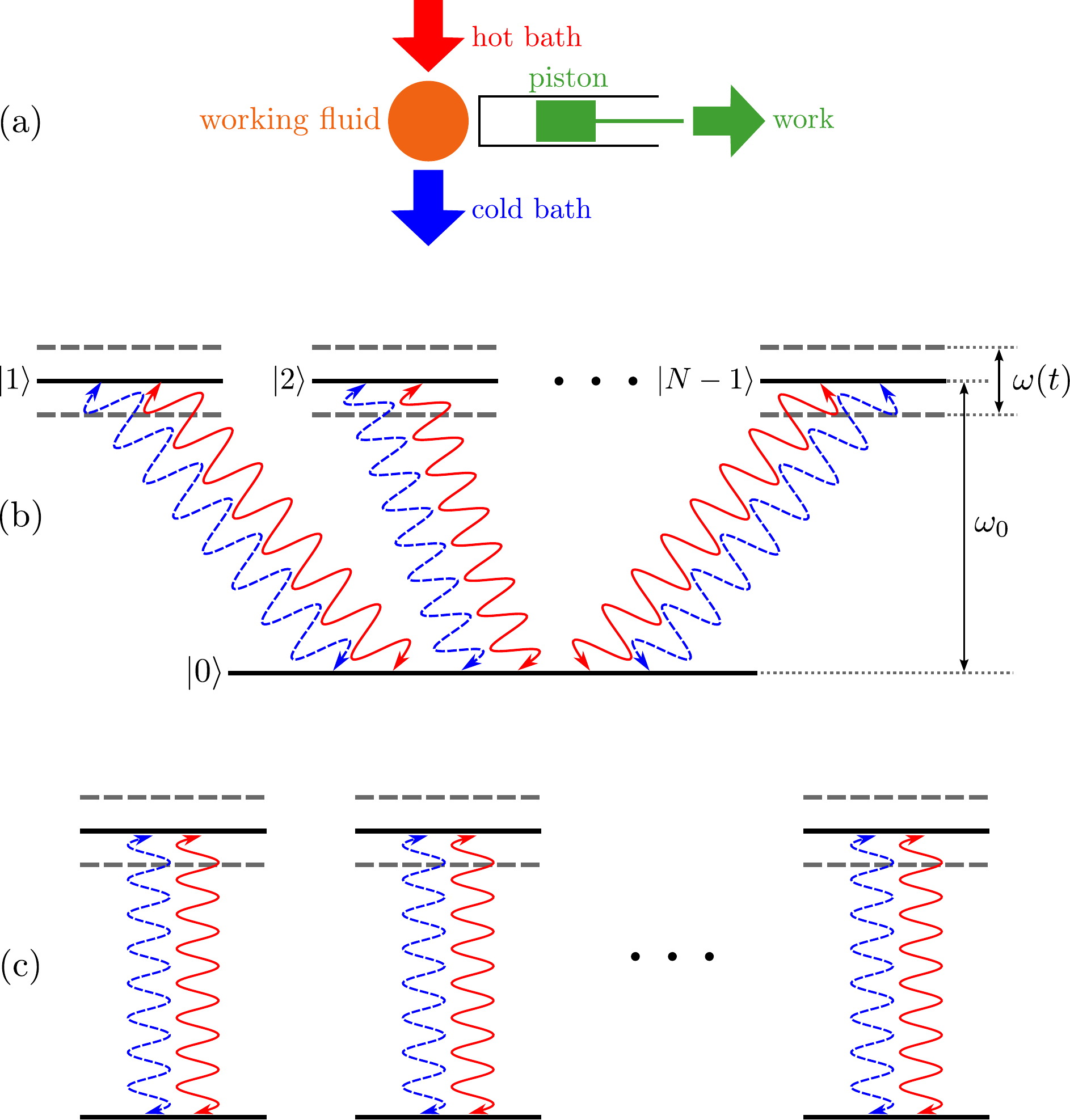}
  \caption{(Color online) (a) Sketch of a heat machine. In a reciprocating-cycle implementation the cold and hot baths as well as the piston are alternately coupled to the working fluid during specific strokes, whereas in continuous-cycle operation the system interacts with both baths as well as the piston at all times. The sketch shows the heat machine operating as an engine. In the refrigeration mode all energy flows (arrows) have to be reversed. (b) A quantum heat machine based on an $N$-level system with ($N-1$)-fold excited-state degeneracy that constantly interacts with a cold (dashed blue) and a hot (solid red) heat bath. The transition frequency is periodically modulated to allow for power extraction (heat engine) or supply (refrigerator). (c) An analogous quantum heat machine based on an ensemble of two-level atoms.}\label{fig_system}
\end{figure}

\par

We consider a generalized $V$-type $N$-level system consisting of $N-1$ degenerate excited states $\ket{1},\dots,\ket{N-1}$ and a common ground state $\ket{0}$. The transition frequency between the excited states and the ground state is denoted by $\omega_0$. This system is assumed to be coupled (alternately or constantly) to cold and hot heat baths and periodically-driven (frequency-modulated) by a piston (see Fig.~\ref{fig_system}). Energy can flow from and to the baths via absorption and (stimulated and spontaneous) emission, respectively, of thermal bath quanta.

\par

The system-bath interaction Hamiltonian is assumed to be of a generalized dipole-coupling form~\cite{breuerbook} that is expressed in the rotating-wave approximation as~\cite{gelbwaser2014power}
\begin{equation}\label{eq_H_SB}
  H_\mathrm{SB}=\sum_{j=1}^{N-1}\sum_{i\in\{\indexc,\indexh\}}\left(\splus^j\otimes\mathbf{d}_j\cdot\mathbf{B}_i+\sminus^j\otimes\mathbf{d}_j^*\cdot\mathbf{B}_i^\dagger\right),
\end{equation}
in terms of the excitation and de-excitation Pauli operators 
\begin{subequations}\label{eq_splus}
  \begin{align}
    \splus^j&\coloneq\ketbra{j}{0}\\
    \sminus^j&\coloneq\ketbra{0}{j},
  \end{align}
\end{subequations}
the (possibly complex) dipole moments for the $j$th transition $\mathbf{d}_j$, and the coupling operators of the cold (hot) heat bath $\mathbf{B}_\indexc$ ($\mathbf{B}_\indexh$). We adopt the notation 
\begin{equation}\label{eq_def_alpha}
  \alpha_j\coloneq|\mathbf{d}_j|/|\mathbf{d}_1|,
\end{equation}
where the strength of the largest transition dipole $|\mathbf{d}_1|$ serves as reference (i.e., $\alpha_j\leq 1$). We further define the dipole alignment matrix
\begin{equation}\label{eq_def_pij}
  \mathfrak{p}_{ij}e^{i\varphi_{ij}}\coloneq\frac{\mathbf{d}_{i}^*\cdot\mathbf{d}_{j}}{|\mathbf{d}_{i}||\mathbf{d}_{j}|}\stackrel{\mathbf{d}_{i},\mathbf{d}_{j}\in\mathbb{R}^3}{\equiv} \cos\measuredangle(\mathbf{d}_{i},\mathbf{d}_{j}),
\end{equation}
where $\mathfrak{p}_{ij}\in[0,1]$, which encodes the directional configuration of the $N-1$ dipoles. This matrix is Hermitian, as $\mathfrak{p}_{ij}=\mathfrak{p}_{ji}$ and $\varphi_{ij}=-\varphi_{ji}$. In what follows we will for brevity denote dipoles $\mathbf{d}_i$ and $\mathbf{d}_j$ with $\mathfrak{p}_{ij}=1$ as being parallel even if they differ by a phase ($\varphi_{ij}\neq 0$).

\par

Prior to discussing in detail a possible setup of such a heat machine, we infer from the system-bath interaction Hamiltonian~\eqref{eq_H_SB} some general results concerning the thermodynamic r\^ole of coherences in the working medium.

\subsection{Collective-states basis}

Owing to the degeneracy of the excited states of the system Hamiltonian
\begin{equation}\label{eq_introduction_H_S}
  H_\mathrm{S}=\hbar\omega_0\sum_{j=1}^{N-1}\splus^j\sminus^j,
\end{equation}
one is free to choose any (rotated) basis within this excited manifold, as the rotated states will still be energy eigenstates of the Hamiltonian~\eqref{eq_introduction_H_S}. Assuming that all dipoles are parallel and (for simplicity) of equal strength and real, we then transform the system-bath Hamiltonian~\eqref{eq_H_SB} to the basis consisting of the ground state $\ket{0}$, the collective ``bright'' state~\cite{ficek2004simulating}
\begin{equation}\label{eq_introduction_psib}
  \ket\psib\coloneq\frac{1}{\sqrt{N-1}}\sum_{j=1}^{N-1}\ket{j},
\end{equation}
and $N-2$ rotated excited states that are orthogonal to the latter. The system-bath interaction Hamiltonian~\eqref{eq_H_SB} then adopts the form
\begin{equation}\label{eq_introduction_H_SB_parallel}
  H_\mathrm{SB}^\mathrm{parallel}=\sum_{i\in\{\indexc,\indexh\}}\sqrt{N-1}\left(\ketbra{\psib}{0}\otimes\mathbf{d}_1\cdot\mathbf{B}_i+\mathrm{H.c.}\right).
\end{equation}
Formally, this Hamiltonian describes a \emph{single two-level system} formed by $\ket{0}$ and $\ket{\psib}$, whose interaction with the baths has a dipole moment \emph{enhanced} by a factor of $\sqrt{N-1}$, which is responsible for superradiance~\cite{dicke1954coherence,gross1982superradiance}. The remaining $N-2$ states, which are orthogonal to $\ket{0}$ and $\ket\psib$, are not accessible by this Hamiltonian: They are \emph{dark states} with respect to the (dipolar) system-bath interaction. Consequently, if the $N$-level system with parallel dipoles is initially prepared in one of these dark states, it does not exchange heat with the baths, and its state remains invariant under the action of the system-bath interaction Hamiltonian. We therefore anticipate the result (elaborated further on), that the steady-state solution for parallel dipoles strongly depends on the overlap of the initial state $\rho(0)$ with these dark states, i.e., on the initial value $\Pidew$ of the dark-state projector.

\par

If, however, the initial state of the multilevel system is non-dark, then the rate of quanta exchange of the \emph{effective} TLS (formed by $\ket\psib$ and $\ket{0}$) with the baths, $\gamma_\mathrm{b}$, is enhanced by a factor of $N-1$ compared to the spontaneous-emission rate $\gamma_1$ of a TLS consisting of the states $\ket{0}$ and $\ket{1}$. This can be deduced from the (dissipative) master-equation description for the reduced density matrix of the system, based on the interaction Hamiltonian~\eqref{eq_introduction_H_SB_parallel}: The spontaneous-emission (decay) rate scales with the square of the transition-dipole moment~\cite{carmichaelbook} and therefore $\gamma_\mathrm{b}=(N-1)\gamma_1$ with $\gamma_1\propto|\mathbf{d}_1|^2$ (Sec.~\ref{subsec_parallel}).

\subsection{Power and efficiency bounds}\label{sec_introduction_efficiency}

Let us consider the implications of the collective enhancement for the heat currents, i.e., the rate of energy exchange between the system and the baths, and the power of the heat machine. The heat currents must be proportional to the rate of quanta absorption and emission of quanta from and to the baths, respectively~\cite{alicki1979quantum,kosloff1984quantum,boukobza2006thermodynamics}. This proportionality should be independent of the actual implementation of the heat machine, i.e., whether both baths are continuously coupled to the system or only one at a time in a reciprocating cycle. 

\par

Due to the enhancement of the decay rate for parallel dipoles in a bright state, the ``cold'' ($\Jc$) and the ``hot'' ($\Jh$) heat currents $J_i=\dot{Q}_i$, $Q_i$ being the heat exchange with the $i$th bath, will be equally enhanced compared to their two-level counterparts, as
\begin{subequations}\label{eq_introduction_currents_tls}
  \begin{align}
    \Jc&=(N-1)\Jc^\mathrm{TLS}\\
    \Jh&=(N-1)\Jh^\mathrm{TLS}.
  \end{align}
\end{subequations}
The first law of thermodynamics (energy conservation)~\cite{alicki1979quantum,kosloff2013quantum,gelbwaser2013minimal} then implies that the power is enhanced by the same factor,
\begin{equation}\label{eq_introduction_power}
  \dot{W}=-(\Jc+\Jh)=(N-1)\dot{W}^\mathrm{TLS}.
\end{equation}
Equations~\eqref{eq_introduction_currents_tls} and~\eqref{eq_introduction_power} imply that although an enhancement of the output power is expected, the efficiency of the multilevel heat engine, which is defined by the ratio of the extracted power to the invested ``hot'' current~\cite{schwablbook},
\begin{equation}\label{eq_introduction_eta}
  \eta\coloneq\frac{|\dot{W}|}{\Jh}=\frac{|\dot{W}^\mathrm{TLS}|}{\Jh^\mathrm{TLS}}\equiv 1-\frac{|\Jc^\mathrm{TLS}|}{\Jh^\mathrm{TLS}},
\end{equation}
remains the \emph{same} as for a TLS-based heat engine where coherence is absent. Notably, the Carnot bound~\cite{schwablbook} is adhered to, by virtue of Eq.~\eqref{eq_introduction_power}, Eq.~\eqref{eq_introduction_eta} and the second-law condition~\cite{spohn1978entropy}
\begin{equation}\label{eq_introduction_second_law}
  \frac{\Jc}{\Tc}+\frac{\Jh}{\Th}\leq 0.
\end{equation}

\par

The assumptions in the foregoing discussion were the degeneracy of the excited states as well as the dipolar-coupling form of $H_\mathrm{SB}$ between the system and both thermal baths. Any increase of the efficiency compared to Eq.~\eqref{eq_introduction_eta} would require different enhancement of $\Jc$ and $\Jh$, such that the enhancement $c_\indexh$ of the ``hot'' current $\Jh=c_\indexh\Jh^\mathrm{TLS}$ exceeds its ``cold'' current counterpart $c_\indexc$, i.e., $c_\indexh>c_\indexc$. This may only be possible for different coupling Hamiltonians between the system and the two thermal baths, $H_\mathrm{SB}^\mathrm{c}$ and $H_\mathrm{SB}^\mathrm{h}$, unlike Eq.~\eqref{eq_H_SB}. However, the Carnot bound, which is a corollary of the first and second laws~\cite{schwablbook}, is upheld regardless of these assumptions since even if such enhancement can be achieved, the second law restricts the ratio of the current enhancement factors $c_\indexc$ and $c_\indexh$ to satisfy [\cf Eq.~\eqref{eq_introduction_second_law}]
\begin{equation}
 c_\indexc\frac{\Jc^\mathrm{TLS}}{\Tc}+c_\indexh\frac{\Jh^\mathrm{TLS}}{\Th}\stackrel{!}{\leq}0.
\end{equation}

\par

\subsection{R\^ole of coherences}

If all dipoles are parallel, coherences in the working fluid (medium) will build up in the bare (energy) basis according to the Hamiltonian~\eqref{eq_introduction_H_SB_parallel} once the system is in contact with heat baths at finite temperature, since the bright state $\ket\psib$ [see Eq.~\eqref{eq_introduction_psib}] is a coherent superposition of the bare excited states. \emph{Initial} coherences may either suppress or enhance the machine performance: They can bring the heat machine to a complete standstill if they correspond to the set of dark states $\{\ket{\psid^i}\}$ since the Hamiltonian~\eqref{eq_introduction_H_SB_parallel} then yields $H_\mathrm{SB}\ket{\psid^i}=0$, so that the system is then decoupled from the baths. Conversely, the Hamiltonian~\eqref{eq_introduction_H_SB_parallel} will enhance both the heat currents and the power by a factor of $N-1$ due to the superradiant (collective-decay) effect if either $\ket\psib$ or the ground state $\ket{0}$ is initially prepared. Yet, even if the system has initially no coherence, it will evolve via the Hamiltonian~\eqref{eq_introduction_H_SB_parallel} to a thermal mixture of $\ket{0}$ and $\ket{\psib}$ that has persistent coherences. Thus, steady-state coherences (embodied by the bright state) are formed in the bare energy basis as a result of thermalization, but the entire system does not thermalize, i.e., an $N$-level Gibbs state does not arise because of the availability of dark states (even if they are not populated) (see Sec.~\ref{subsec_parallel}).

\par

The foregoing discussion of the thermodynamic r\^ole of coherences in the working medium holds for the special case of all transition-dipole vectors in the Hamiltonian~\eqref{eq_H_SB} being parallel. In what follows, we, however, wish to explore the possibility of power enhancement for misaligned (non-parallel) dipoles, or any combination of aligned and misaligned dipoles. Since steady-state coherences correspond to (destructive or constructive) interference of the dipoles, they can only arise if at least some of the dipoles are aligned. Only then may dark states exist. Without the presence of such dark states, a thermal (Gibbs) steady state of the working fluid is expected.

\par

As we show in Sec.~\ref{sec_dicke}, these arguments also hold for an ensemble of $N$ two-level atoms in a suitable geometry, which realizes the Dicke model~\cite{dicke1954coherence,gross1982superradiance} (Sec.~\ref{sec_multiatom_realization}). In that case the effective system-bath interaction Hamiltonian cannot be written in the form of Eq.~\eqref{eq_introduction_H_SB_parallel}. Instead, the ensemble can then be mapped onto a collective spin-$N/2$ system and a multitude of dark states. The heat currents will no longer adhere to a simple form as in Eqs.~\eqref{eq_introduction_currents_tls}, but the efficiency of the Dicke heat machine will still be the same as of the TLS-based machine.

\section{Floquet expansion of the master equation in continuous cycles}\label{sec_system}

In the remainder of this article these issues are investigated for a continuous-cycle heat machine, wherein the ($N-1$)-fold degenerate system is constantly coupled to the two thermal baths and to a periodically modulating ``piston'' that allows for work extraction (in the case of an engine) or supply (in the case of a refrigerator or heat pump), respectively (see Fig.~\ref{fig_system}). 

\par

In our model the ``piston'' effects are described by a synchronous periodic modulation of all the upper states~\cite{gelbwaser2013minimal,alicki2014quantum,gelbwaser2014power},
\begin{equation}\label{eq_H_S}
  H_\mathrm{S}(t)=\hbar[\omega_0+\omega(t)]\sum_{j=1}^{N-1}\splus^j\sminus^j,
\end{equation}
where $\omega(t+\frac{2\pi}{\Omega})=\omega(t)$, $\Omega$ being the modulation rate. Such a transition-energy modulation may for example be induced by a varying magnetic field (via the Zeeman effect) or by an alternating electric field (via the Stark effect).

\par

The master equation~\cite{carmichaelbook} for the reduced density operator $\rho$ in the interaction picture is
\begin{equation}\label{eq_master_general}
  \dot\rho=\mathcal{L}\rho,
\end{equation}
where the Liouvillian superoperator $\mathcal{L}$ is of the Lindblad--Gorini--Kossakowski--Sudarshan (LGKS) form~\cite{gorini1976completely,lindblad1976generators}. As shown in~\cite{alicki2012periodically,gelbwaser2013minimal} this Liouvillian can be decomposed into ``sub-bath'' Liouvillians $\mathcal{L}_i^q$ describing the interaction of the system with the $i$th bath ($i\in\{\indexc,\indexh\}$) evaluated at the $q$th harmonic sideband ($q\in\mathbb{Z}$) of the unperturbed transition frequency $\omega_0$, induced by the modulation rate $\Omega$. These ``sub-bath'' (sideband) contributions are a consequence of the Floquet theorem for the solution of linear differential equations with periodic coefficients. The master equation~\eqref{eq_master_general} then adopts the \emph{additive} form
\begin{subequations}\label{eq_master_L}
  \begin{equation}\label{eq_master}
    \dot\rho=\sum_{q\in\mathbb{Z}}\sum_{i=\{\indexc,\indexh\}}\mathcal{L}_i^q\rho
  \end{equation}
  with the ``sub-bath'' Liouvillians (generalizing Ref.~\cite{gelbwaser2014power})
  \begin{widetext}
    \begin{multline}\label{eq_L}
      \mathcal{L}_i^q\rho=\frac{1}{2}P(q)G_i(\omega_0+q\Omega)\sum_{j=1}^{N-1}\left[\alpha_j^2\mathcal{D}\left(\sminus^j,\splus^j\right)+\sum_{\substack{j^\prime\neq j}}\mathfrak{p}_{jj^\prime}e^{i\varphi_{jj^\prime}}\alpha_j\alpha_{j^\prime}\mathcal{D}\left(\sminus^j,\splus^{j^\prime}\right)\right]+\\
      +\frac{1}{2}P(q)G_i(-\omega_0-q\Omega)\sum_{j=1}^{N-1}\left[\alpha_j^2\mathcal{D}\left(\splus^j,\sminus^j\right)+\sum_{\substack{j^\prime\neq j}}\mathfrak{p}_{jj^\prime}e^{-i\varphi_{jj^\prime}}\alpha_j\alpha_{j^\prime}\mathcal{D}\left(\splus^j,\sminus^{j^\prime}\right)\right].
    \end{multline}
  \end{widetext}
\end{subequations}

\par

Here the dissipator is defined as $\mathcal{D}(a,b)\coloneq 2a\rho b-ba\rho-\rho ba$ for arbitrary system operators $a,b$. The terms $\mathcal{D}\left(\sminus^j,\splus^j\right)$ and $\mathcal{D}\left(\splus^j,\sminus^j\right)$ in Eq.~\eqref{eq_L} describe (spontaneous and stimulated) emission into and absorption from the bath, respectively, involving a single transition dipole $\mathbf{d}_j$ and weighted by $\alpha_j^2$ [\cf Eq.~\eqref{eq_def_alpha}]. By virtue of sharing a \emph{common} ground state $\ket{0}$, these contributions to the Liouvillian amount to an \emph{indirect population transfer} between the different excited states via this common ground state.
\par
By contrast, the cross-terms $\mathcal{D}\left(\sminus^j,\splus^{j^\prime}\right)$ and $\mathcal{D}\left(\splus^j,\sminus^{j^\prime}\right)$ in Eq.~\eqref{eq_L}, weighted by $\alpha_j\alpha_{j^\prime}$, describe \emph{correlated} absorption and emission involving two different transitions, i.e., quanta exchange between excited states $\ket{j}$ and $\ket{j^\prime}$ via the common ground state $\ket{0}$, $\ket{j}\rightarrow\ket{0}\rightarrow\ket{j^\prime}$. These bath-mediated interactions between the excited states---which result in dynamical coherences---are largest when the two dipole moments involved, $\mathbf{d}_j$ and $\mathbf{d}_{j^\prime}$, are parallel up to a phase factor ($\mathfrak{p}_{jj^\prime}=1$). For orthogonal dipole orientations ($\mathfrak{p}_{jj^\prime}=0\ \forall j\neq j^\prime$) these cross-correlations vanish and no coherences build up. In three-dimensional space, this completely orthogonal configuration can only be realized in a three- or four-level system.

\par

The rates of all decay and absorption processes are determined by the respective prefactors of the dissipators in the Liouvillian~\eqref{eq_L}. Here $P(q)$ are the Floquet coefficients~\cite{alicki2012periodically} determining the weight of the $q$th harmonic sideband (with normalization $\sum_{q\in\mathbb{Z}}P(q)=1$) and $G_i(\omega)$ is the response spectrum of the $i$th bath evaluated at frequency $\omega$. These spectra fulfill the Kubo--Martin--Schwinger (KMS) detailed-balance condition~\cite{breuerbook}
\begin{equation}\label{eq_kms}
  G_i(-\omega)=e^{-\beta_i\hbar\omega}G_i(\omega). 
\end{equation}
For a bosonic bath, the coupling strengths explicitly read 
\begin{equation}
G_i(\omega)=\gamma_i(\omega)\left(\bar{n}_i(\omega)+1\right),
\end{equation}
with $\bar{n}_i(\omega)\coloneq(e^{\beta_i\hbar\omega}-1)^{-1}$ denoting the number of thermal quanta at inverse temperature $\beta_i=1/\kB T_i$ and $\gamma_i(\omega)$ being the frequency-dependent transition rate induced by the $i$th bath.

\section{Three-level system}\label{sec_ehrenfest}

In Sec.~\ref{sec_system} we have introduced the full model, invoking couplings to two heat baths and periodic modulation of $N-1$ excited states. These ingredients are necessary for the operation of the system as a heat machine. However, all the ``sub-bath'' Liouvillians $\mathcal{L}_i^q$ [Eq.~\eqref{eq_L}] in the master equation~\eqref{eq_master} contain the \emph{same} dissipators $\mathcal{D}$ and only differ by their respective prefactors. Owing to this additive structure, in this section we first solve the master equation for a three-level system interacting with a single bath at inverse temperature $\beta=1/\kB T$ and a static (unmodulated) transition frequency ($q=0$, $P(0)=1$), and analyze its steady-state solution. This solution will serve to obtain the ``global'' solution for two baths involving a modulated transition frequency by simple changes of the coefficients (i.e., the transition rates). The two-bath ``global'' solution allowing for the modulation-induced harmonic sidebands, which is required for the computation of the heat currents, will be presented in Sec.~\ref{sec_heat_currents}.

\subsection{Steady-state solution for degenerate excited states}\label{sec_threelevels_degenerate}

In order to obtain some insight into the dynamics, we transform the master equation~\eqref{eq_master} into a set of Ehrenfest equations of motion for operator expectation values. These equations can subsequently be cast into an inhomogeneous linear ordinary differential equation (ODE) system
\begin{equation}\label{eq_ode}
  \dot{\mathbf{x}}=\mathcal{A}\mathbf{x}+\mathbf{b}
\end{equation}
for the density matrix elements. In the case $N=3$ this vector is
\begin{equation}
  \mathbf{x}\coloneq(\rho_{21},\rho_{12},\rho_{00},\rho_{22})^T.
\end{equation}
The corresponding coefficient matrix $\mathcal{A}$ and the inhomogeneity $\mathbf{b}$ are presented in Appendix~\ref{app_threelevels}.

\par

This linear system does not describe the entire density matrix, as it does not contain the coherences $\rho_{01}$ and $\rho_{02}$ between the ground and the excited states. The reason is that those matrix elements are completely decoupled from $\mathbf{x}$ and obey the independent (homogeneous) differential equation
\begin{equation}\label{eq_ode_y}
  \dot{\mathbf{y}}=\mathcal{B}\mathbf{y}
\end{equation}
with
\begin{equation}
  \mathbf{y}\coloneq(\rho_{10},\rho_{01},\rho_{20},\rho_{02})^T
\end{equation}
and the coefficient matrix $\mathcal{B}$ given in Appendix~\ref{app_threelevels}. All eigenvalues of the matrix $\mathcal{B}$ are negative. Consequently, the coherences between the ground and excited states are damped out regardless of their initial values and the steady-state solution of Eq.~\eqref{eq_ode_y} has no coherences, i.e., $\mathbf{y}^\mathrm{ss}=\mathbf{0}$. 

\par

Let us now revisit the ODE~\eqref{eq_ode} for $\mathbf{x}$. The uniqueness of its steady-state solution depends on the determinant of the coefficient matrix $\mathcal{A}$,
\begin{multline}\label{eq_detA}
  \det(\mathcal{A})=\left[\frac{1}{2}G(\omega_0)\right]^44\alpha^2\left(1+2e^{-\beta\hbar\omega_0}\right)\times\\\times\left(1+\alpha^2\right)^2\left(1-\mathfrak{p}^2\right).
\end{multline}
For aligned dipoles ($\mathfrak{p}\coloneq\mathfrak{p}_{12}=1$) we find $\det(\mathcal{A})=0$, corresponding to a singularity of the coefficient matrix~\eqref{eq_odes_a}. In this regime, multiple steady-state solutions of Eq.~\eqref{eq_ode} may exist (depending on the initial conditions), each satisfying the linear system of equations
\begin{equation}
  \mathcal{A}\mathbf{x}^\mathrm{ss}=-\mathbf{b}.
\end{equation}

\par

The general steady-state solution of the linear ODE~\eqref{eq_ode} (Eq.~\eqref{eq_ss_threelevels} in Appendix~\ref{app_threelevels}) is compatible with the one found in Refs.~\cite{agarwal2001quantum,ficek2002quantum}. While the steady state is diagonal and unique for non-aligned dipoles ($\mathfrak{p}\neq 1$), it depends on the initial conditions (via an integral of motion) and yields persistent coherences $\rho_{21}^\mathrm{ss}=\left(\rho_{21}^\mathrm{ss}\right)^*$ in the aligned case ($\mathfrak{p}=1$). The integral of motion agrees with a previous finding (involving a zero-temperature bath with~\cite{kozlov2006inducing} and without~\cite{ficek2002quantum} incoherent external driving, respectively). Such initial-condition dependent steady-state solutions have also been found for degenerate $\Lambda$-systems~\cite{berman2005spontaneously}.

\par

\begin{figure}
  \centering
  \includegraphics[width=\columnwidth]{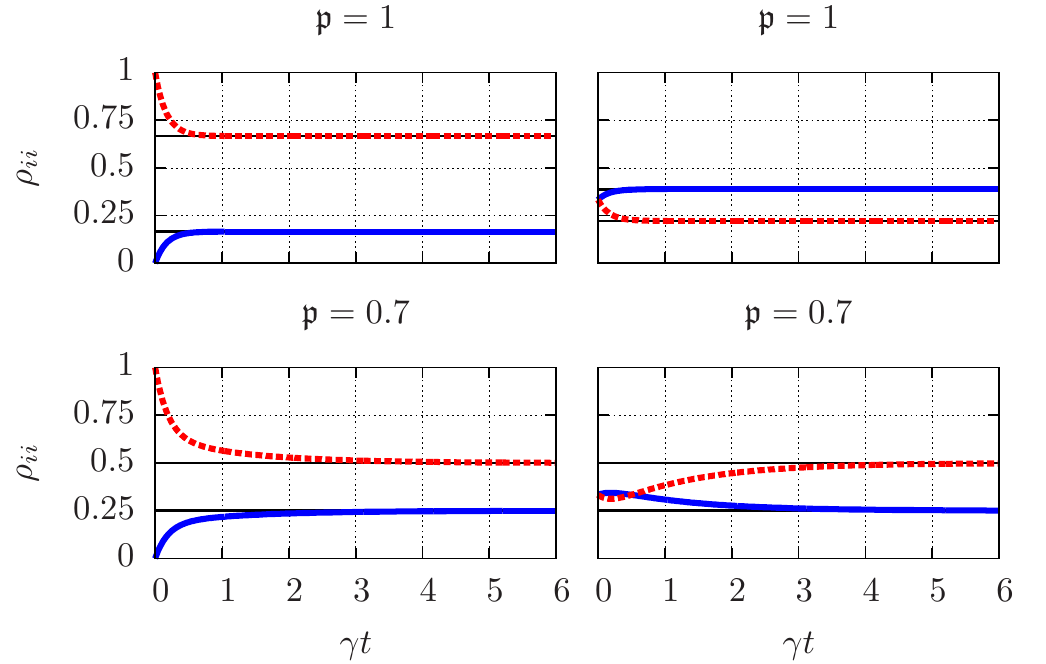}
  \caption{(Color online) Time evolution of the populations (solid blue: $\rho_{22}$, dashed red: $\rho_{00}$) for aligned ($\mathfrak{p}=1$) and misaligned ($\mathfrak{p}=0.7$) dipoles and different initial conditions $\rho(0)=\proj{0}$ (left) and $\rho(0)=\proj{\psi}$ with $\ket\psi=\left(\ket{0}+\ket{1}-\ket{2}\right)/\sqrt{3}$ (right)
obtained by numerically integrating the master equation~\eqref{eq_master}. The horizontal solid black lines are the analytic steady-state solution~\eqref{eq_ss_threelevels} of the ODE~\eqref{eq_ode}. Parameters: $e^{-\beta\hbar\omega_0}=\frac{1}{2}$, $\alpha=1$ and $G(\omega_0)=\gamma(\bar{n}+1)\equiv2\gamma$.}\label{fig_timeevolution_populations}
\end{figure}

\begin{figure}
  \centering
  \includegraphics[width=\columnwidth]{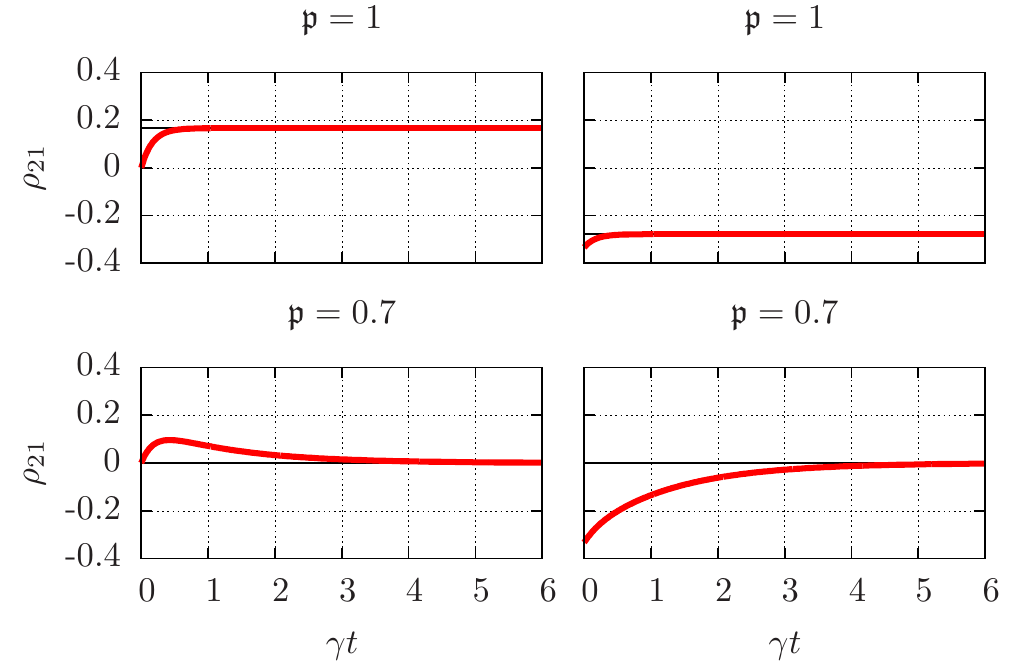}
  \caption{(Color online) Same as Fig.~\ref{fig_timeevolution_populations} for the coherence $\rho_{21}$. In the aligned case ($\mathfrak{p}=1$) the coherence persists in steady state, whereas it is merely a transient effect in the misaligned case ($\mathfrak{p}=0.7$).}\label{fig_timeevolution_coherences}
\end{figure}

The time evolution of the system towards the steady state~\eqref{eq_ss_threelevels} is illustrated in Figs.~\ref{fig_timeevolution_populations} and~\ref{fig_timeevolution_coherences}, where we have numerically integrated the master equation~\eqref{eq_master} and compared its steady state to the analytic steady-state solution~\eqref{eq_ss_threelevels}. The time evolution of the ODE~\eqref{eq_ode} gives the same result. The dependence of the steady-state solution on the initial conditions for aligned dipoles ($\mathfrak{p}=1)$ can clearly be seen. Coherences between the excited states persist only in this regime and are just a transient effect for any misalignment. Notably, as seen from Eqs.~\eqref{eq_ss_threelevels}, the steady states for orthogonal dipoles (without transient coherences) and any other misalignment coincide. Consequently, dynamical correlations between the upper states [induced by the cross-terms in the Liouvillian~\eqref{eq_L}] are important at finite times but not in the long-time limit. In particular, this implies that coherences cannot play any \emph{thermodynamic} r\^ole in misaligned dipole transitions in the steady-state operation of the heat machine.

\par

\subsection{Non-degeneracy effects}

The singularity of the coefficient matrix $\mathcal{A}$ for aligned dipoles only arises for perfectly degenerate levels. If a detuning $\Delta$ between the two excited states is introduced into the coefficient matrix~\eqref{eq_odes_a}, the determinant becomes proportional to $\left[(1+\alpha^2)^2(1-\mathfrak{p}^2)+\Delta^2/G^2(\omega_0)\right]$ in the Schr\"odinger picture [as a generalization of Eq.~\eqref{eq_detA}]. Amongst the solutions of $\det(\mathcal{A})=0$, only the case $\mathfrak{p}=1$ \emph{and} $\Delta=0$ lies within the domain of existence of $\mathfrak{p}$, which is $[0,1]$. In the interaction picture (in which the master equation~\eqref{eq_master} is presented) the coefficients then become time dependent through the phase factors $\mathfrak{p}e^{\pm i\varphi}\mapsto\mathfrak{p}e^{\pm i\varphi}e^{\pm i\Delta t}$, and the steady state is no longer determined by the determinant of the coefficient matrix. In the secular approximation, these time-dependent phase factors, and hence all cross-terms in the Liouvillian~\eqref{eq_L}, vanish upon coarse-graining the time evolution~\cite{breuerbook}. The steady-state solution is then given by the canonical (Gibbs) distribution (diagonal in the energy basis) $\rho_{ii}^\mathrm{ss}=e^{-\beta\hbar\omega_i}\rho_{00}^\mathrm{ss}$ for $1\leq i\leq N-1$. Hence, we must assume \emph{degenerate} levels in order to study possible implications of quantum coherences on the thermodynamic properties of a \emph{steady-state} quantum heat machine.

\section{Steady-state solution: Dependence of thermalization on the dipole orientations}\label{sec_ss}

In the previous section we have shown that in the simplest case ($N=3$), aligned dipoles are dynamically distinct from all other possible geometric configurations (\cf also Ref.~\cite{gelbwaser2014power}). In this section we show that this distinction arises from the existence of a dark state (anticipated in Sec.~\ref{sec_H_SB}), which hinders full thermalization under certain initial conditions if the dipoles are all aligned. We expect similar behavior for the general case of $N-1$ dipole transitions. In the following we investigate the steady-state solution for arbitrary dipole orientations, taking into account that as $N$ grows, so does the dipole configuration space. To this end we first consider the extreme, simple cases of (i) no dipoles being pairwise parallel and (ii) all dipoles being parallel. The general solution for arbitrary dipole orientations can then be derived in a transparent way from these extreme cases.

\subsection{No pairwise parallel dipoles}\label{sec_ss_pneq1}

If no dipoles are parallel, the master equation~\eqref{eq_master} (for a single bath at inverse temperature $\beta$ and without modulation, $q=0$) possesses a unique thermalized, diagonal, steady state
\begin{subequations}\label{eq_ss_nlevels_pneq1}
  \begin{align}
    \rho_{ii}^\mathrm{ss}&=\frac{1}{N-1+e^{\beta\hbar\omega_0}}\equiv e^{-\beta\hbar\omega_0}\rho_{00}^\mathrm{ss}\quad i\in[1,\dots,N-1]\\
    \rho_{00}^\mathrm{ss}&=\frac{1}{1+(N-1)e^{-\beta\hbar\omega_0}}.
  \end{align}
\end{subequations}
The straightforward proof is given in Appendix~\ref{app_proof_pneq1}. Here too, the steady state is independent of the actual orientation of the dipoles (as long as they are not aligned)---but the time evolution leading to this steady state depends on their orientation.

\par

Owing to their degeneracy, any superposition of excited states is again an energy eigenstate. This means that the density matrix~\eqref{eq_ss_nlevels_pneq1} is not only diagonal in the ``bare'' basis $\{\ket{0},\ket{1},\dots,\ket{N-1}\}$, but also in any basis $\{\ket{0},\ket{1^\prime},\dots,\ket{(N-1)^\prime}\}$ formed by the ground state and the rotated excited states.

\subsection{Parallel dipoles}\label{subsec_parallel}

Let us now consider the opposite case of all dipoles being parallel (which has been previously discussed in Sec.~\ref{sec_H_SB}). The steady-state solution in this case may not be unique due to the singularity of the coefficient matrix, which is linked to the existence of dark and bright states.

\par

When all dipoles are aligned with the reference dipole $\mathbf{d}_{1}$, i.e., $\mathbf{d}_{j}=\alpha_je^{i\varphi_{1j}}\mathbf{d}_{1}$, the interaction Hamiltonian~\eqref{eq_H_SB} between the system and the bath can be recast into the generic form [\cf also Eq.~\eqref{eq_introduction_H_SB_parallel}]
\begin{equation}\label{eq_H_SB_parallel}
  \left.H_\mathrm{SB}\right|_{\mathfrak{p}_{ij}=1}=\sqrt{\sum_{j=1}^{N-1}\alpha_j^2}\left(\splusbar\otimes \mathbf{d}_{1}\cdot\mathbf{B} + \sminusbar\otimes \mathbf{d}_{1}^*\cdot\mathbf{B}^\dagger\right).
\end{equation}
This operator is reported in the basis spanned by the following states: (i) the ground state $\ket{0}$, (ii) the bright state [see also Ref.~\cite{ficek2004simulating} and Eq.~\eqref{eq_introduction_psib}]
\begin{equation}\label{eq_psib}
  \ket{\psib}\coloneq\frac{1}{\sqrt{\sum_{j=1}^{N-1}\alpha_j^2}}\sum_{j=1}^{N-1}\alpha_je^{i\varphi_{1j}}\ket{j},
\end{equation}
and (iii) $N-2$ dark states $\left\{\ket{\psid^j}\right\}$, obtained by the Gram--Schmidt orthogonalization procedure.
\par
The Hamiltonian~\eqref{eq_H_SB_parallel} \emph{formally} looks like the coupling Hamiltonian of a single two-level system interacting with the environment. However, this operator acts on an $N$-dimensional Hilbert space, even if only two levels (the ground and the bright states) are explicitly involved, via the Pauli excitation and de-excitation operators [\cf also Eq.~\eqref{eq_introduction_H_SB_parallel}]
\begin{subequations}\label{eq_pauli_bright_states}
  \begin{align}
    \splusbar&\coloneq\ketbra{\psib}{0}\\
    \sminusbar&\coloneq\ketbra{0}{\psib}.
  \end{align}
\end{subequations}
The associated transition-dipole vector is scaled by a factor of $\sqrt{\sum_{j=1}^{N-1}\alpha_j^2}$, which for equal transition strengths ($\alpha_j=1\,\forall j$) amounts to $\sqrt{N-1}$. The diagonal steady-state solution of the master equation following from the interaction Hamiltonian~\eqref{eq_H_SB_parallel} has the form (\cf Appendix~\ref{app_proof_p1})
\begin{subequations}\label{eq_ss_nlevels_p1}
  \begin{align}
    \rho_\mathrm{bb}^\mathrm{ss}&=\frac{1}{1+e^{\beta\hbar\omega_0}}\Big[1-\Pidew\Big]\equiv e^{-\beta\hbar\omega_0}\rho_{00}^\mathrm{ss}\\
    \rho_{00}^\mathrm{ss}&=\frac{1}{1+e^{-\beta\hbar\omega_0}}\Big[1-\Pidew\Big]\\
    \rho_{kk}^\mathrm{ss}&=\bkew{\phi_k}{\rho(0)}{\phi_k}\quad\text{ for }k=1,\dots,N-2.
  \end{align}
\end{subequations}
Here we have defined the projector
\begin{equation}
  \Pid\coloneq\sum_{j=1}^{N-2}\proj{\psid^j}\equiv\mathbbm{1}-\proj{\psib}-\proj{0}
\end{equation}
onto the dark-state subspace, the dark-state populations $\rho_{kk}^\mathrm{ss}$, and the spectral decomposition of the part of the initial state lying within the dark-state subspace,
\begin{equation}
 \rho_\mathrm{d}(0)=\sum_{i,j=1}^{N-2}q_{ij}(0)\ketbra{\psid^i}{\psid^j}=\sum_{k=1}^{N-2}p_k(0)\proj{\phi_k}.
\end{equation}
This density matrix is not normalized to unity but to $\Tr\rho_{\mathrm{d}}(0)=\Pidew\equiv 1-\bkew{\psib}{\rho(0)}{\psib}-\bkew{0}{\rho(0)}{0}$. The states $\ket{\phi_k}$ associated to the dark-state populations $\rho_{kk}^\mathrm{ss}$ are thus the eigenvectors of $\rho_\mathrm{d}(0)$. If $\rho_\mathrm{d}(0)=0$, any basis of the dark subspace may be chosen instead.

\par

The dark-state populations $\rho_{kk}^\mathrm{ss}$ in the steady-state solution~\eqref{eq_ss_nlevels_p1} are the same as their initial values, since these states do not interact with the environment. The bright- and ground-state components (which do interact with the environment) eventually thermalize, their Boltzmann factor $e^{-\beta\hbar\omega_0}$ being determined by the bath temperature. However, the steady-state thermalization is only partial if the initial state has an overlap with the dark-state subspace. We therefore refer to the factor $1-\Pidew$ as the \emph{thermalization capability} of the initial state $\rho(0)$.

\par

The foregoing discussion has shown that whilst the interpretation of the initial-condition-dependent terms in the non-diagonal density matrix~\eqref{eq_ss_threelevels} may be obscure, their physical meaning is clearly revealed in the diagonal steady-state form~\eqref{eq_ss_nlevels_p1}, which has a distinct non-Gibbs character.

\subsection{Arbitrary dipole orientations}

Based on the two limiting cases considered above, we expect the dynamics to be different for aligned (parallel) and misaligned transition dipoles.
\par
\begin{figure}
  \centering
  \includegraphics[width=\columnwidth]{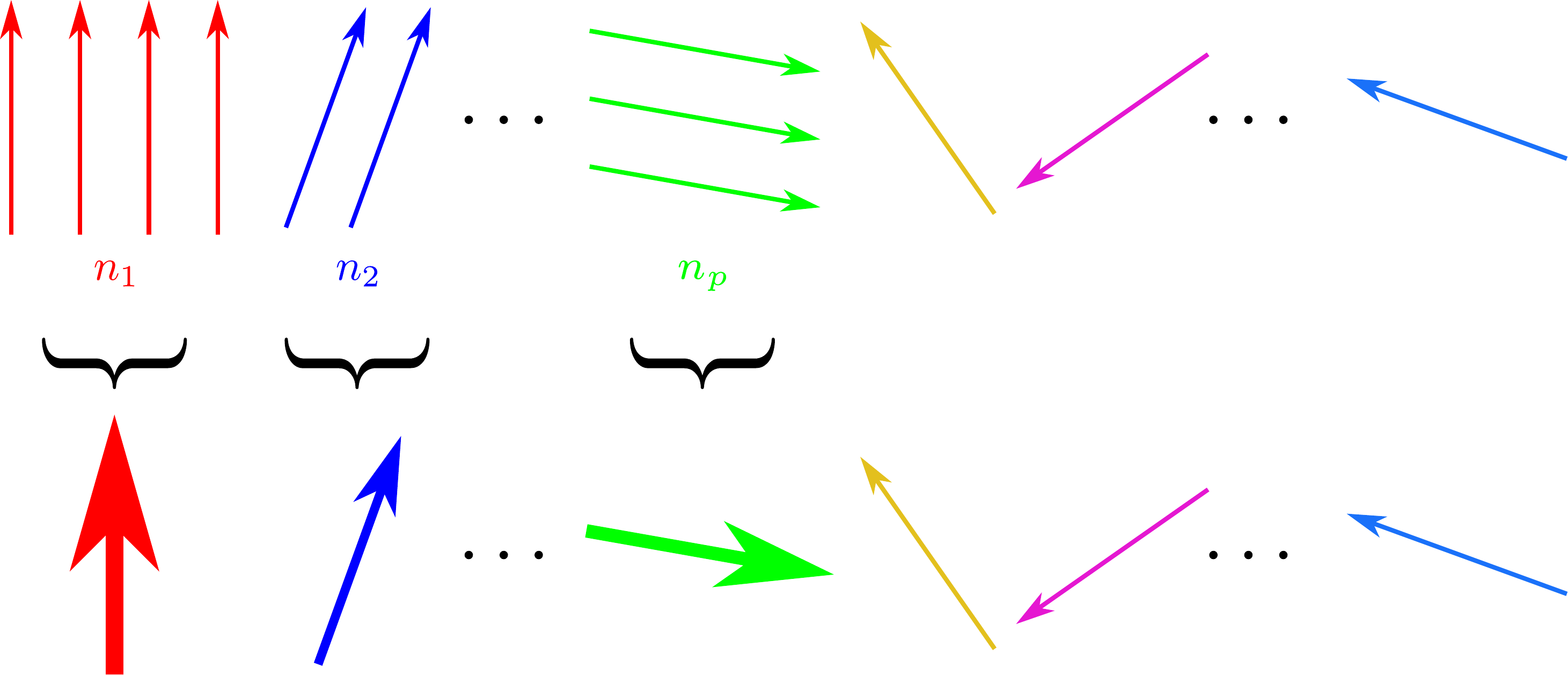}
  \caption{(Color online) A possible configuration exhibiting a domain structure of parallel dipole transitions. Each group (domain) of aligned transitions (upper arrows) is equivalent to a single (enhanced) dipole transition (lower arrows).}\label{fig_dipoles}
\end{figure}

We therefore group all aligned transition dipoles (up to a phase factor) into ``domains'': The first $n_1$ aligned dipoles are grouped in domain $1$, the next $n_2$ parallel ones are grouped in domain $2$ and so forth. Altogether, there are $p$ ``domains'', which contain
\begin{equation}
  N_p\coloneq\sum_{j=1}^{p} n_j
\end{equation}
transition dipoles. The remaining $N-1-N_p$ transition dipoles are non-parallel to any other dipole (see Fig.~\ref{fig_dipoles}).

\par
\par
\par
\begin{figure}
  \centering
  \includegraphics[width=\columnwidth]{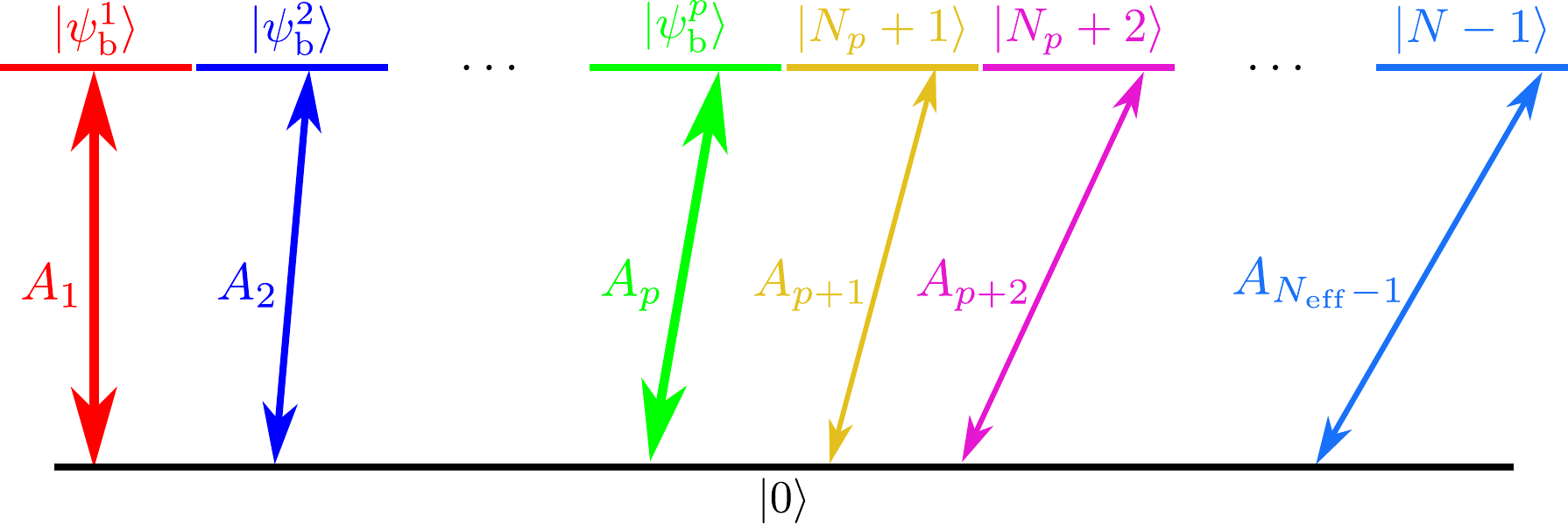}
  \caption{(Color online) Effective $\Neff$-level scheme corresponding to the $N$-level system in Fig.~\ref{fig_dipoles}. The transitions associated to a bright state (i.e., to a domain of parallel dipoles) are enhanced compared to their bare-state counterparts. Within this effective system no transition dipoles are parallel to each other.}\label{fig_dipoles_effective_level_scheme}
\end{figure}
\par

As detailed in Appendix~\ref{app_H_SB_arbitrary_geometry}, the $N$-level system with arbitrary dipoles can be mapped onto an $\Neff$-level system with \emph{non-parallel dipoles}, where
\begin{equation}\label{eq_neff}
  \Neff\coloneq p+N-N_p.
\end{equation}
Upon introducing the new coupling constants
\begin{equation}
  A_m\coloneq
  \begin{cases}
    \sqrt{\sum_{j=n_{m-1}+1}^{n_m}\alpha_{j}^2} & \text{ for } m\leq p \\
    \alpha_{N_p+(m-p)} & \text{ for } p+1\leq m\leq\Neff-1
  \end{cases}
  ,
\end{equation}
the interaction Hamiltonian adopts the form
\begin{equation}\label{eq_H_SB_arbitrary}
  H_\mathrm{SB}=\sum_{m=1}^{\Neff-1} A_m|\mathbf{d}_1|\left(\splustilde^m\otimes \mathbf{e}_m\cdot\mathbf{B} + \sminustilde^m\otimes \mathbf{e}_m^*\cdot\mathbf{B}^\dagger\right).
\end{equation}
This is simply the Hamiltonian~\eqref{eq_H_SB} for $\Neff$ levels with modified coupling strengths and \emph{no} pairwise parallel dipoles. The new Pauli operators $\splustilde^m$ are [\cf Eqs.~\eqref{eq_splus} and~\eqref{eq_splusbar}]
\begin{equation}
  \splustilde^m\coloneq
  \begin{cases}
    \splusbar^m\equiv\ketbra{\psib^m}{0}&m\leq p\\
    &\\
    \parbox[t]{0.35\columnwidth}{$\splus^{N_p+(m-p)}$\\\mbox{$\equiv\ketbra{N_p+(m-p)}{0}$}}&p+1\leq m\leq \Neff-1
  \end{cases}
  .
\end{equation}
\par
The mapping from $N$ levels with arbitrary transition dipoles to $\Neff$ non-aligned dipoles (\cf Figs.~\ref{fig_dipoles} and~\ref{fig_dipoles_effective_level_scheme}) now assigns a weight $A_j$ to each transition, according to the number of \emph{thermalization pathways} forming each domain of parallel dipoles. Clearly, the completely misaligned ($\Neff=N$) and aligned ($\Neff=2$) cases discussed in Secs.~\ref{sec_ss_pneq1} and~\ref{subsec_parallel} above are included in this general scheme.

\par

The Liouvillian associated with the Hamiltonian~\eqref{eq_H_SB_arbitrary} is thus the same as for the non-aligned system obtained upon replacing $N\mapsto\Neff$ and $\alpha_j\mapsto A_j$. The steady-state solution of the master equation then reads
\begin{subequations}\label{eq_ss_nlevels_general}
  \begin{align}
    \rho_{00}^\mathrm{ss}&=\frac{1}{1+(\Neff-1)e^{-\beta\hbar\omega_0}}\left[1-\Pidew\right]\label{eq_ss_nlevels_general_ground_state}\\
    \rho_{ii}^\mathrm{ss}&=
    \begin{cases}
      e^{-\beta\hbar\omega_0}\rho_{00}^\mathrm{ss}&i=1,\dots,\Neff-1\\
      \bkew{\phi_i}{\rho(0)}{\phi_i}&i=\Neff+1,\dots,N
    \end{cases}
    .
  \end{align}
\end{subequations}
Here again the $\{\ket{\phi_i}\}$ are the eigenvectors of $\rho_{\mathrm{d}}(0)$, i.e., the part of the initial density matrix lying within any of the $p$ dark-state subspaces. The projector $\Pid$ onto the union of all dark-space subspaces has now the form
\begin{align}
  \Pid&=\sum_{j=1}^{p}\sum_{k=1}^{n_k-1}\proj{\psid^{j,k}}\equiv\\
  &\equiv\mathbbm{1}-\sum_{j=1}^{p}\proj{\psib^{j}}-\sum_{j=N_p+1}^{N-1}\proj{j}-\proj{0}.
\end{align}
The proof of Eqs.~\eqref{eq_ss_nlevels_general} is analogous to the respective proofs of the steady-state solutions~\eqref{eq_ss_nlevels_pneq1} and~\eqref{eq_ss_nlevels_p1} for non-aligned and aligned dipoles.

\par

Whilst the steady-state density matrix~\eqref{eq_ss_nlevels_general} is diagonal in the new basis associated with the Hamiltonian~\eqref{eq_H_SB_arbitrary} (and hence does not contain any coherences), off-diagonal elements do appear in ${\tilde\rho}^\mathrm{ss}$, the representation of the density matrix $\rho^\mathrm{ss}$ in the energy (bare-state) basis [\cf also Eq.~\eqref{eq_ss_threelevels}].

\par

It is important to note that the coherences with the largest modulus, $\sum_{i,j=1 (i\neq j)}^{N-1}|{\tilde\rho}_{ij}^\mathrm{ss}|$, arise for a \emph{dark initial state}, since the exchange between the excited states [described by Eqs.~\eqref{eq_master_L}] implies that the ground state is at least partly (thermally) populated in steady state [\cf Eq.~\eqref{eq_ss_nlevels_general_ground_state}]. This ground-state population inevitably reduces the excited-state population and hence the coherences between excited states. Equivalently, the steady-state density matrix can only describe a completely population-inverted state iff the initial state is dark. Yet, dark states are a severe hindrance to the operation of a quantum heat machine. Hence, the magnitude of steady-state coherences cannot be a criterion for possible power enhancement in a multilevel quantum heat engine, as shown below in detail.

\section{Heat currents and power}\label{sec_heat_currents}

\subsection{Heat currents}

The general steady-state solution~\eqref{eq_ss_nlevels_general} of the master equation~\eqref{eq_master} is the key ingredient for computing the heat flows (currents) between the system and the cold and hot baths. Upon invoking the dynamical version of the second law of thermodynamics and the von~Neumann entropy, the heat currents associated to the $i$th bath are found to be~\cite{alicki2012periodically,gelbwaser2013minimal,kosloff2013quantum} 
\begin{equation}
  J_i=\sum_{q\in\mathbb{Z}}J_i^q
\end{equation}
in terms of the harmonic sideband contributions
\begin{equation}\label{eq_J}
  J_i^q\coloneq-\frac{1}{\beta_i}\Tr\left[(\mathcal{L}_i^q\rho^\mathrm{ss})\ln(\rho_i^q)\right].
\end{equation}
Here $\rho^\mathrm{ss}$ is the \emph{global} steady-state solution of the master equation~\eqref{eq_master} satisfying 
\begin{equation}
  \mathcal{L}\rho^\mathrm{ss}=0,
\end{equation}
and $\rho_i^q$ is the \emph{local} solution for the system interacting with a single ($i$th) bath evaluated at the $q$th harmonic sideband, i.e., the steady-state solution of the Liouvillian~\eqref{eq_L}
\begin{equation}
  \mathcal{L}_i^q\rho_i^q=0. 
\end{equation}
The heat currents~\eqref{eq_J} are based on a global two-bath solution, as required to avoid violation of the second law~\cite{levy2014local}.

\par

The steady-state solution of the entire master equation~\eqref{eq_master} can be obtained from Eq.~\eqref{eq_ss_nlevels_general} upon adapting the transition rates to a two-bath situation. Owing to the KMS relation~\eqref{eq_kms}, this amounts to replacing the real inverse temperature $\beta$ by an \emph{effective} inverse temperature $\betaeff$, which is defined by the Boltzmann factor 
\begin{equation}\label{eq_betaeff}
  e^{-\betaeff\hbar\omega_0}\coloneq\frac{\sum_{q\in\mathbb{Z}}\sum_{i\in\{\indexc,\indexh\}} P(q)G_i(-\omega_0-q\Omega)}{\sum_{q\in\mathbb{Z}}\sum_{i\in\{\indexc,\indexh\}} P(q)G_i(\omega_0+q\Omega)}.
\end{equation}
This effective temperature can be controlled (modified) by engineering the modulation type [and thus the Floquet coefficients $P(q)$] or the cold- and hot-bath response spectra [$\Gc(\omega_0+q\Omega)$ and $\Gh(\omega_0+q\Omega)$] at the corresponding harmonic sidebands $q\in\mathbb{Z}$, respectively. Likewise, the local steady-state solution $\rho_i^q$ is obtained from Eq.~\eqref{eq_ss_nlevels_general} upon replacing
\begin{subequations}
  \begin{align}
    \beta&\mapsto\beta_i\\
    \omega_0&\mapsto\omega_0+q\Omega.
  \end{align}
\end{subequations}

\par

Inserting $\rho^\mathrm{ss}$ and $\rho_i^q$ into expression~\eqref{eq_J}, we find the heat currents, whose explicit form is presented in Appendix~\ref{app_heat_currents_power}. Their comparison to the previously found result for a TLS~\cite{gelbwaser2013minimal} yields the heat current ratio
\begin{equation}\label{eq_heat_currents_general_tls}
  \frac{J_i}{J_i^\mathrm{TLS}}=\left(\sum_{j=1}^{N-1}\alpha_j^2\right)\Big[1-\Pidew\Big]\frac{1+e^{-\betaeff\hbar\omega_0}}{1+(\Neff-1)e^{-\betaeff\hbar\omega_0}}.
\end{equation}
Here the first factor on the right-hand side is the enhancement that stems from the $N-1$ available thermalization pathways. It amounts to $N-1$ if all transitions are of the same strength ($\alpha_j=1\,\forall j$). The second factor measures the thermalization capability of the initial state (i.e., the overlap of the initial state with the non-dark states, viz.\ the part of the initial state amenable to thermalization). This factor is only relevant for the case of at least two dipoles being parallel (for other configurations there are no dark states). The last factor strongly depends on the modulation type and the bath properties via the inverse effective temperature $\betaeff$ defined in Eq.~\eqref{eq_betaeff}, but also on the effective number of misaligned dipoles $\Neff$ [\cf Eq.~\eqref{eq_neff}]. This last factor becomes unity for $\Neff=2$. Such an \emph{effective} two-level system is obtained for an $N$-level system with all transition dipoles being aligned [\cf Eqs.~\eqref{eq_ss_nlevels_p1}].

\par

Remarkably, for any number of misaligned dipoles the product of the last two factors in Eq.~\eqref{eq_heat_currents_general_tls} is determined by the ratio of the respective ground-state populations, yielding
\begin{equation}\label{eq_heat_currents_general_tls_rho00}
   \frac{J_i}{J_i^\mathrm{TLS}}=\left(\sum_{j=1}^{N-1}\alpha_j^2\right)\frac{\rho_{00}^\mathrm{ss}}{\rho_{00}^\mathrm{TLS}}\leq N-1.
\end{equation}

\par

It follows from Eq.~\eqref{eq_heat_currents_general_tls_rho00} that the largest possible enhancement factor is bounded by
\begin{equation}
  \sum_{j=1}^{N-1}\alpha_j^2\leq N-1,  
\end{equation}
the equality sign corresponding to all transitions having the same strength. This maximal enhancement corresponds to the combined heat currents from $N-1$ \emph{independent} two-level heat machines.

\par

\begin{figure}
  \centering
  \includegraphics[width=\columnwidth]{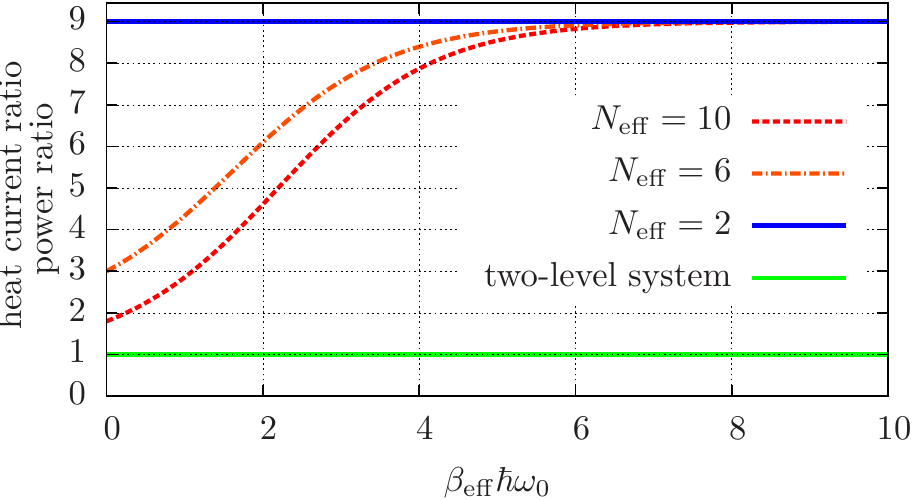}
  \caption{(Color online) Heat current ratio $J_i/J_i^\mathrm{TLS}$ [Eq.~\eqref{eq_heat_currents_general_tls}] for a ten-level system assuming no initial overlap with any dark states for (i) $\Neff=N=10$ (no parallel dipoles), (ii) $\Neff=6$ (some dipoles parallel) and (iii) $\Neff=2$ (all nine dipoles parallel). Alignment is only beneficial at high effective temperatures (small $\betaeff\hbar\omega_0$). For all dipoles being parallel, amplification is independent on the bath temperatures. Due to Eq.~\eqref{eq_power_general_tls} the same holds for the power ratio.}\label{fig_currents_nlevels}
\end{figure}

The most advantageous initial condition is a state orthogonal to the dark subspace, e.g., $\rho(0)=\proj{0}$. In that case, as is apparent from Fig.~\ref{fig_currents_nlevels}, at low effective temperatures all dipole orientations yield the same (maximum) enhancement by a factor of $N-1$ (assuming all dipoles are equally strong). This can be understood from the population ratio in Eq.~\eqref{eq_heat_currents_general_tls_rho00}: At low temperatures the ground-state population is close to one, independent of the dipole configuration. 
\par

In the opposite case of high effective temperatures, configurations with more parallel transitions are more favorable. This can again be explained based on the ground-state population ratio in Eq.~\eqref{eq_heat_currents_general_tls_rho00}. Assuming full thermalization, the ground state is the more populated the fewer levels are available. Hence, for $N$ levels the smallest $\Neff$ are the most beneficial at such temperatures. For the $10$-level system shown in Fig.~\ref{fig_currents_nlevels}, the ground-state population in the high-temperature limit ($\betaeff\rightarrow0$) is [according to Eqs.~\eqref{eq_ss_nlevels_general}] $\rho_{00}^\mathrm{ss}=1/\Neff$, compared with $\rho_{00}^\mathrm{TLS}=1/2$, so that Eq.~\eqref{eq_heat_currents_general_tls_rho00} yields $\lim_{\betaeff\rightarrow0}J_i/J_i^\mathrm{TLS}=9\times2/\Neff$.

\par
\par

\subsection{Power and efficiency}

The first law (energy conservation) $\dot{W}=-(\Jh+\Jc)$ relates the heat currents to the power. Whilst the explicit form of the power is again reported in Appendix~\ref{app_heat_currents_power}, we here compare it to its TLS counterpart,
\begin{equation}\label{eq_power_general_tls}
  \frac{\dot{W}}{\dot{W}^\mathrm{TLS}}=\left(\sum_{j=1}^{N-1}\alpha_j^2\right)\frac{\rho_{00}^\mathrm{ss}}{\rho_{00}^\mathrm{TLS}}\equiv\frac{J_i}{J_i^\mathrm{TLS}}\leq N-1.
\end{equation}
This is the same ratio as in Eq.~\eqref{eq_heat_currents_general_tls_rho00} for the heat currents and therefore Fig.~\ref{fig_currents_nlevels} also holds for the power ratio. We have thus come to a central conclusion: An $N$-level quantum heat machine can \emph{never} outperform $N-1$ independent two-level heat machines.

\par

In contrast to the power, the efficiency
\begin{equation}
  \eta=\frac{-\dot{W}}{\Jh}
\end{equation}
(when the heat machine acts as an engine) and the coefficient of performance
\begin{equation}
  \mathrm{COP}=\frac{\Jc}{\dot{W}}
\end{equation}
(when the heat machine acts as a refrigerator), respectively, are \emph{not} affected by the presence of degenerate upper states, regardless of the dipole orientation. Notably, the Carnot bound is not violated, based on the results previously obtained for the TLS heat machine~\cite{gelbwaser2013minimal}, as discussed in Sec.~\ref{sec_introduction_efficiency}.

\subsection{Power-dependence on modulation rate}

The result~\eqref{eq_power_general} for the power output of the periodically-driven continuous multilevel quantum heat machine only differs by the extra factor $\Neff-1$ in the denominator from its counterpart for a two-level machine discussed in Ref.~\cite{gelbwaser2013minimal}. It was shown there that depending on the modulation frequency $\Omega$ the heat machine can be operated either as an engine ($\dot{W}<0$ for $\Omega<\Omegacrit$) or as a refrigerator/heat pump ($\dot{W}>0$ for $\Omega>\Omegacrit$), with $\Omegacrit$ being some critical frequency.

\par

This behavior can be understood in terms of the frequency dependence of the bath populations $n_\indexc(\omega)$ and $n_\indexh(\omega)$. At the TLS resonance frequency, they fulfill $n_\indexh(\omega_0)>n_\indexc(\omega_0)$. The periodic driving introduces harmonic sidebands, that shift the frequencies at which the TLS couples to the baths [\cf also the Liouvillian~\eqref{eq_L}]. When only two sidebands $\omega_0\pm\Omega$ contribute and the cold and hot bath spectra do not overlap, the ``natural'' direction of the heat flows is reversed provided that $n_\indexh(\omega_0+\Omega)<n_\indexc(\omega_0-\Omega)$~\cite{kolar2012quantum}. This explains the occurrence of a ``critical'' frequency $\Omegacrit$, defined by $n_\indexh(\omega_0+\Omegacrit)=n_\indexc(\omega_0-\Omegacrit)$, at which the heat flows (and hence also the power) vanish. This is the point at which Carnot efficiency is reached~\cite{gelbwaser2013minimal}. Similar scenarios occur for example for ultracold atoms in cavities~\cite{ritsch2013cold}, where, depending on the frequency of the cavity-pump laser, the atoms are either cooled or heated.

\subsection{Power-dependence on initial conditions for aligned dipoles}

\begin{figure}
  \centering
  \includegraphics[width=\columnwidth]{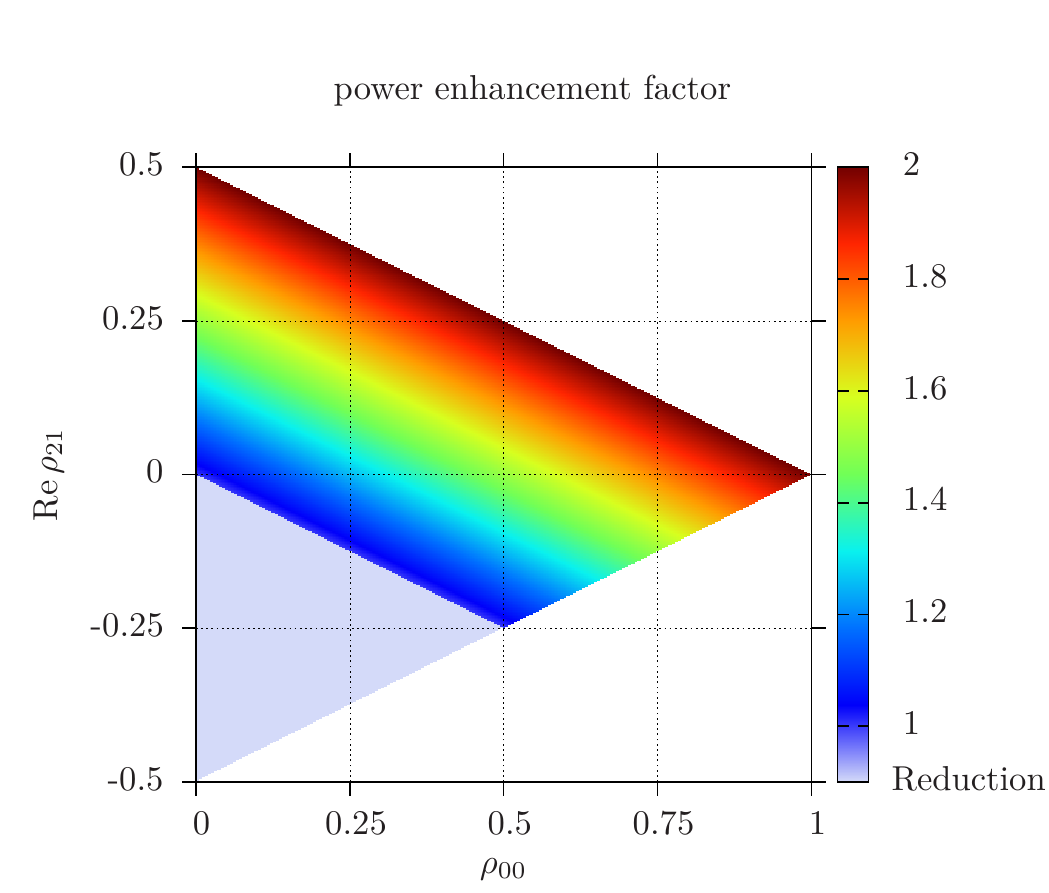}
  \caption{(Color online) Power enhancement factor~\eqref{eq_power_general_tls_parallel_three_level_system} for a three-level system with respect to a TLS as a function of the ground-state population and the coherence for parallel equal dipole moments. The optimal amplification (by a factor of two) is realized on the upper edge of the colored area, i.e., for the maximally allowed coherence (and correct phase) once $\rho_{00}$ is fixed [Eq.~\eqref{eq_initial_state_coherence_rho00_fixed}]. The lower left shaded triangle corresponds to a reduction of the power relative to a TLS, up to complete suppression in its lower left corner (the point corresponding to the dark state, which has the maximally allowed value for the modulus of the coherence but the opposite phase). For anti-parallel ($\varphi=\pi$) dipoles the diagram would be vertically flipped around the $\realt\rho_{21}=0$ axis.}\label{fig_power_enhancement}
\end{figure}

As we have seen, the power enhancement of the $N$-level heat machine compared to a single two-level heat machine strongly depends on the initial conditions if some of the transition-dipole vectors are aligned [Eq.~\eqref{eq_power_general}]. We here illustrate this dependence for a three-level ($N=3$) system, which allows for a simple graphical representation. We shall now specify its advantageous initial conditions. Once the ground-state population $\rho_{00}(0)$ is fixed, the advantageous states, amenable to full thermalization, are those having the largest possible modulus of the coherence and the appropriate phase. These ``optimal'' coherences and populations must satisfy (\cf Sec.~\ref{sec_threelevels_degenerate})
\begin{subequations}
  \begin{align}
    \rho_{21}(0)&=\frac{\alpha e^{i\varphi}}{1+\alpha^2}\left[1-\rho_{00}(0)\right]\label{eq_initial_state_coherence_rho00_fixed}\\
    \rho_{11}(0)&=\frac{1}{1+\alpha^2}\left[1-\rho_{00}(0)\right]\\
    \rho_{22}(0)&=\frac{\alpha^2}{1+\alpha^2}\left[1-\rho_{00}(0)\right].
  \end{align}
\end{subequations}
These conditions may be satisfied by a pure state $\rho(0)=\proj{\psi(0)}$, where
\begin{equation}
  \ket{\psi(0)}=\sqrt{\rho_{00}(0)}\ket{0}+\sqrt{1-\rho_{00}(0)}e^{i\Phi}\ket{\psib},
\end{equation}
with an arbitrary phase $\Phi$. Alternatively, these conditions hold (since the coherences between the ground- and excited states do not matter) for the incoherent mixture 
\begin{equation}
  \rho(0)=\rho_{00}(0)\proj{0}+\left[1-\rho_{00}(0)\right]\proj{\psib}
\end{equation}
in the new basis containing the bright state [\cf Eq.~\eqref{eq_psib}]
\begin{equation}
\ket\psib=\frac{1}{\sqrt{1+\alpha^2}}\left(\ket{1}+\alpha e^{i\varphi}\ket{2}\right).  
\end{equation}
Any deviation from this ``optimal'' initial condition results in a reduced thermalization capability and therefore in a reduced power output. The power-enhancement factor for two equal parallel dipoles ($\alpha=1$ and $\varphi=0$) is then
\begin{equation}\label{eq_power_general_tls_parallel_three_level_system}
  \frac{\dot{W}}{\dot{W}^\mathrm{TLS}}=2\Big[1-\Pidew\Big].
\end{equation}
Figure~\ref{fig_power_enhancement} clearly reveals that the initial state has to be carefully prepared in order to exhibit the maximal possible power boost.

\par
\par

\section{Dicke-system performance}\label{sec_dicke}

\par
\begin{figure}
  \centering
  \includegraphics[width=\columnwidth]{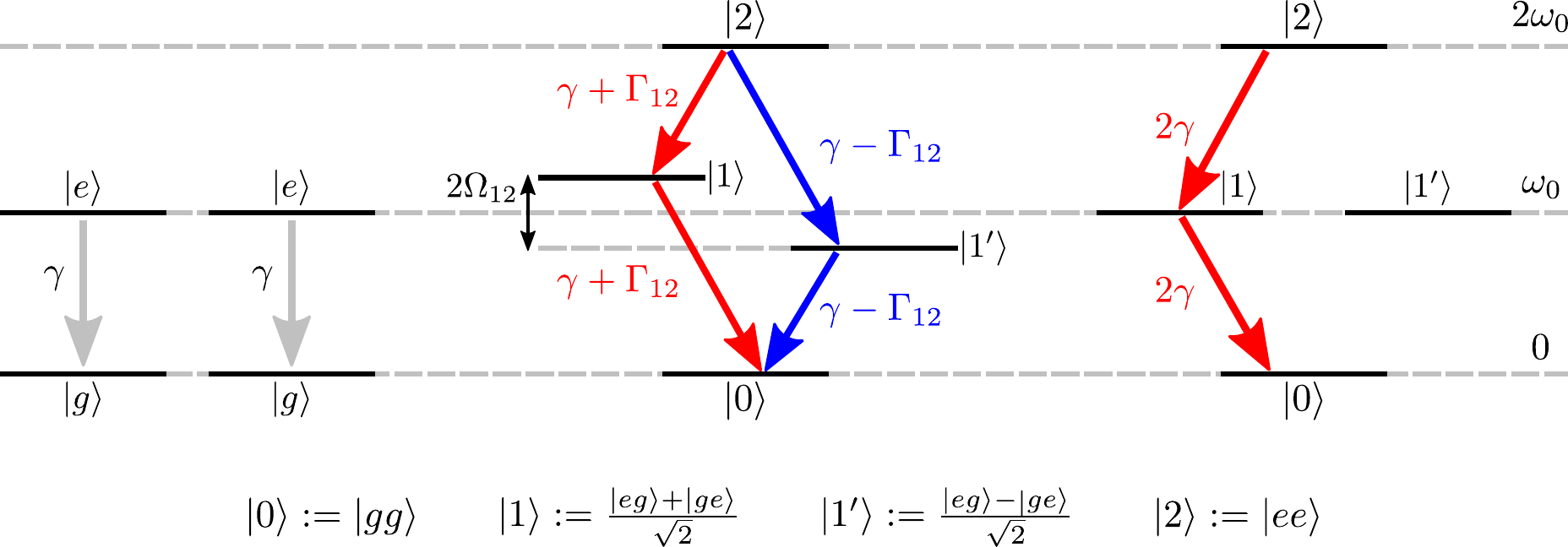}
  \caption{(Color online) The Dicke system exemplified for $N=2$. The two atoms interacting with the same environment (left plot) are mapped onto a non-degenerate four-level system consisting of collective states (middle plot). The two states carrying one excitation are superradiant ($\ket{1}$) and subradiant ($\ket{1^\prime}$), respectively. If the dipole-dipole interaction between the atoms vanishes ($\Omega_{12}=0$) and their collective decay rates are maximized ($\Gamma_{12}=\gamma$), the system is mapped onto an effective three-level system with equidistant levels (equivalent to a spin-$1$ system plus one dark state) (right plot).}\label{fig_dicke}
\end{figure}
\par

Let us now consider an ensemble of $N$ two-level atoms interacting with the same bath. The bath mediates the dipole-dipole interaction as well as non-local cooperative photon exchange, whereby a photon emitted by one atom can be re-absorbed by another~\cite{lehmberg1970radiation,lehmberg1970radiation2}. Quite generally, the dipole-dipole interaction renders the $j$-excitation states non-degenerate~\cite{lehmberg1970radiation2,agarwalbook,ficek1986cooperative}. Whilst some of these collective states give rise to an enhanced decay rate, others decay much slower than their single-atom counterparts (see Fig.~\ref{fig_dicke}), conforming to ``superradiance'' and ``subradiance'', respectively~\cite{agarwalbook,agarwal1970master,lehmberg1970radiation,lehmberg1970radiation2,ficek1986cooperative,prasad2000polarium,scully2006directed,akkermans2008photon,scully2009super,lin2012superradiance,svidzinsky2013quantum}. Hence, in general $N$-atom systems do not obey the Dicke model~\cite{dicke1954coherence}. However, by choosing an appropriate geometry (elaborated on in Sec.~\ref{sec_multiatom_realization}), the dipole-dipole interaction can be made to vanish, while at the same time the subradiant states become completely dark (see Fig.~\ref{fig_dicke}). Only such a configuration can realize the Dicke system---$N$ two-level atoms (equivalent to spin-$1/2$ particles) can then be mapped onto a collective spin-$N/2$ system~\cite{breuerbook}, whose Hilbert space is spanned by $N+1$ (symmetrized) collective states $\ket{j}$ containing $j$ excitations with eigenenergy $j\hbar\omega_0$, where $\omega_0$ denotes the TLS transition frequency, and $2^N-(N+1)$ dark states, which are decoupled from the dynamics (By contrast, in the fully aligned multilevel case $N-2$ states of the total $N$ states are dark). This mapping~\cite{breuerbook,lehmberg1970radiation2} is sketched in Fig.~\ref{fig_dicke} for two atoms.

\par

The dissipative dynamics of the Dicke system is described by the Liouvillian superoperator (for brevity given here for a single bath and without modulation) (see also Ref.~\cite{breuerbook})
\begin{multline}\label{eq_L_dicke}
  \mathcal{L}\rho=N\frac{1}{2}G(\omega_0)\left(2A\rho A^\dagger-A^\dagger A\rho-\rho A^\dagger A\right) \\
  +N\frac{1}{2}G(\omega_0)e^{-\beta\hbar\omega_0}\left(2A^\dagger\rho A-AA^\dagger\rho-\rho AA^\dagger\right),
\end{multline}
with
\begin{equation}
  A\coloneq\frac{1}{\sqrt{N}}\sum_{i=1}^N \sminus^i\equiv\sum_{j=0}^{N-1}\ketbra{j}{j+1}
\end{equation}
and the individual Pauli operators $\sminus^i\coloneq\ketbra{g_i}{e_i}$ for the $i$th two-level atom. The jump operator $\propto \sqrt{N\gamma}A$ describes the emission of one quantum accompanied by de-excitations ($j\rightarrow j-1$) of the non-dark excited states $\ket{j>0}$. The individual emission rates from $\ket{j}$ to $\ket{j-1}$ ($j\in(0,N]$) evaluate to $\gamma_j=\gamma j(N-j+1)$~\cite{breuerbook}. The steady-state solution of the master equation with the Liouvillian~\eqref{eq_L_dicke} is then---in analogy to Eqs.~\eqref{eq_ss_nlevels_general}---the diagonal, partially-thermalized, state
\begin{subequations}\label{eq_ss_dicke}
  \begin{align}
    \rho_{00}^\mathrm{ss}&=\frac{1}{\sum_{j=0}^N e^{-j\beta\hbar\omega_0}}\Big[1-\Pidew\Big]\\
    \rho_{jj}^\mathrm{ss}&=
      \begin{cases}
        e^{-j\beta\hbar\omega_0}\rho_{00}^\mathrm{ss}&j=1,\dots,N\\
        \bkew{\phi_j}{\rho(0)}{\phi_j}&j=N+1,\dots,2^N-1,
      \end{cases}
  \end{align}
\end{subequations}
where we again resort to the projector $\Pid$ onto the dark subspace. Here the states $\ket{\phi_j}$ again denote the eigenstates of the dark part of the initial density matrix. The major differences of Eqs.~\eqref{eq_ss_dicke} to Eqs.~\eqref{eq_ss_nlevels_general} are the Boltzmann factors, which now differ according to the number of excitations in each state $\ket{j}$. As in Eqs.~\eqref{eq_ss_nlevels_general}, coherences between the bare atomic states explicitly appear when transforming the solution~\eqref{eq_ss_dicke} to the $N$-atom product basis.

\par

The same derivation as before for the multilevel case yields the cold and hot heat currents presented in Appendix~\ref{app_heat_currents_dicke}. The ratio of these heat currents~\eqref{eq_dicke_heat_currents} and the associated power to their counterparts generated by $N$ \emph{independent} two-level heat machines [Eq.~\eqref{eq_heat_currents_general} with $N=\Neff=2$] reads
\begin{equation}\label{eq_heat_power_ratio_dicke}
  \frac{J_i}{N J_i^\mathrm{TLS}}\equiv\frac{\dot{W}}{N {\dot{W}}^\mathrm{TLS}}=\left[\sum_{j=0}^{N-1} e^{-j\betaeff\hbar\omega_0}\right]\frac{\rho_{00}^\mathrm{ss}}{\rho_{00}^\mathrm{TLS}},
\end{equation}
where
\begin{equation}
  \frac{\rho_{00}^\mathrm{ss}}{\rho_{00}^\mathrm{TLS}}=\frac{1+e^{-\betaeff\hbar\omega_0}}{\sum_{j=0}^N e^{-j\betaeff\hbar\omega_0}}\Big[1-\Pidew\Big].
\end{equation}

\par

At low effective temperatures, $\betaeff\rightarrow\infty$, at most a single excitation is present in the Dicke system. In this limit, the Dicke ladder for $N-1$ atoms may be mapped onto a degenerate $N$-level system with parallel transition-dipole vectors (which corresponds to an effective two-level system with transition-dipole strength enhanced by $\sqrt{N-1}$). Now, from Fig.~\ref{fig_currents_nlevels} we know that in the low-temperature regime the aligned and non-aligned multilevel systems perform similarly, the latter corresponding to $N-1$ independent TLS. This behavior is obtainable also from Eq.~\eqref{eq_heat_power_ratio_dicke},
\begin{equation}
  \lim_{\betaeff\rightarrow\infty}\frac{\dot{W}}{N {\dot{W}}^\mathrm{TLS}}=\Big[1-\Pidew\Big].
\end{equation}
Hence, at low effective temperatures the $N$-atom Dicke heat machine performs as well as (but not better than) $N$ independent two-level atoms, provided it is amenable to complete thermalization.

\par

In the opposite limit of high effective temperatures, $\betaeff\rightarrow 0$, the cooperative Dicke system yields a power boost,
\begin{equation}
  \lim_{\betaeff\rightarrow0}\frac{\dot{W}}{N {\dot{W}}^\mathrm{TLS}}=\frac{2N}{N+1}\Big[1-\Pidew\Big].
\end{equation}
This ratio has the following meaning: The denominator represents the thermal equipartition of all non-dark state populations, $\Big[1-\Pidew\big]/(N+1)$, whereas $N$ is the bright-state power enhancement and the factor of $2$ stems from the high-temperature TLS ground-state population $1/2$.

\par

\begin{figure}
  \centering
  \includegraphics[width=\columnwidth]{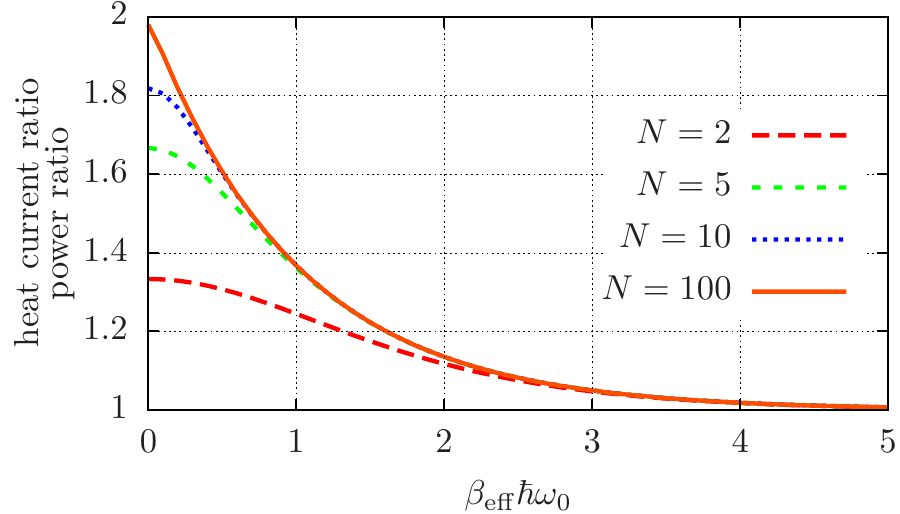}
  \caption{(Color online) Heat currents and power output [Eq.~\eqref{eq_heat_power_ratio_dicke}] for the $N$-atom Dicke model compared to $N$ independent two-level-atom heat machines for different particle numbers. Here we have assumed an optimal initial condition orthogonal to the dark subspace. The maximum enhancement factor is $2$.}\label{fig_currents_dicke}
\end{figure}

Hence, the $N$-atom Dicke heat engine can give at most twice the power of its counterpart consisting of $N$ independent TLS. The full temperature dependence of the power boost for various ensemble sizes is shown in Fig.~\ref{fig_currents_dicke}.

\par
\par

\par

As a consequence of Eq.~\eqref{eq_heat_power_ratio_dicke}, i.e., the identical boost of the two heat currents, the efficiency of the Dicke heat engine and the COP of the Dicke refrigerator, respectively, are the same as their two-level counterparts,
\begin{subequations}
  \begin{align}
    \eta^\mathrm{Dicke}&=\eta^\mathrm{TLS}\\
    \mathrm{COP}^\mathrm{Dicke}&=\mathrm{COP}^\mathrm{TLS}.
  \end{align}
\end{subequations}

\par

Our results on power enhancement in $N$-level systems as well as for $N$ atoms conforming to the Dicke model are summarized in Table~\ref{table}.

\par

\begin{table}
  \centering
  \begin{tabular}{|c|c|c|c|}
    \hline
     & \makecell{maximal power \\ relative to TLS \\ (low $\Teff$)} & \makecell{maximal power \\ relative to TLS \\ (high $\Teff$)} & \makecell{initial state \\ for maximal \\ power} \\
    \hline
    \makecell{$N$-level \\ (non-aligned \\ dipoles)} & $N-1$ & $(N-1)\frac{2}{N}$ & any\\
    \hline
    \makecell{$N$-level \\ (fully aligned \\ dipoles)} & $N-1$ & $N-1$ & non-dark \\
    \hline
    \makecell{$N$-atom \\ Dicke} & $N$ & $\frac{2N^2}{N+1}$ & non-dark \\
    \hline
  \end{tabular}
  \caption{Summary of the (maximal) power enhancement factors and their respective dependencies on the effective temperature [\cf Figs.~\ref{fig_currents_nlevels} and~\ref{fig_currents_dicke}] and the initial condition for the multilevel heat machine and the Dicke machine (compared to a single TLS). The Dicke heat machine shows the best performance.}\label{table}
\end{table}

\section{Realization considerations}\label{sec_realizations}

\subsection{Misaligned dipoles}

Degenerate excited states with non-parallel dipole-transition vectors to the ground state are ubiquitous in atomic systems without hyperfine splitting. For example, the $2p$ manifold of hydrogen consists of the three degenerate states $\ket{n=2,l=1,m=\{0,\pm1\}}$. Their corresponding dipole transition vectors to the ground state are all of equal strength, but orthogonal to each other. This orthogonality, however, does not restrict thermalization (but may not produce the maximal boost at high temperatures, \cf Fig.~\ref{fig_currents_nlevels}), since there is no difference between orthogonal and any other non-aligned configuration in the steady-state heat machine.

\subsection{Aligned dipoles in multilevel atoms}\label{subsec_realizations_parallel}

Multilevel degeneracy combined with perfect alignment (parallelism) of the transition dipoles, which is a prerequisite for coherence effects, is much harder to realize, since it cannot occur in bare atomic systems due to selection rules~\cite{ficek2002quantum}. It is, in general, also not possible to engineer dressed states within the excited manifold that have this property and at the same time adhere to our model: Combining degenerate states with orthogonal dipole moments (as in the $2p$ manifold of hydrogen) creates effective states with non-orthogonal but also non-parallel dipoles~\cite{ficek2002quantum}. 

\par

\begin{figure}
  \centering
  \includegraphics[width=\columnwidth]{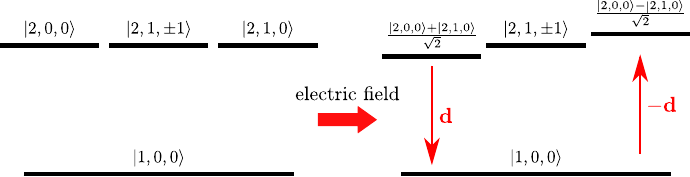}
  \caption{(Color online) Linear Stark effect in hydrogen. The metastable state $\ket{2,0,0}$ is mixed with $\ket{2,1,0}$ to form the dressed states $(\ket{2,0,0}\pm\ket{2,1,0})/\sqrt{2}$ by a static electric field~\cite{schwablbookqm1}; for a weak field these states are quasi-degenerate. Their transition dipoles (red arrows) are anti-parallel, but since $\ket{2,0,0}$ and $\ket{2,1,0}$ have opposite parity, a hydrogen atom in a static electric field behaves as if it possesses a permanent dipole moment~\cite{schwablbookqm1}.}\label{fig_linear_stark_effect}
\end{figure}

A possible realization of alignment may be to superpose a decaying and a non-decaying (metastable) state. Such linear combinations of decaying and metastable states, however, do not possess a definite parity. This means that the dipole operator $\mathbf{D}$ has permanent (static) dipole moments $\bkew{e_i}{\mathbf{D}}{e_i}$, as e.g., formed via the linear Stark effect in hydrogen (see Fig.~\ref{fig_linear_stark_effect}). In the derivation of the master equation~\eqref{eq_master}, however, we have assumed that the ground and excited states have a definite parity and therefore the atom has no permanent dipole moment.

\par
\par

\par

Nevertheless, a permanent dipole moment $\bkew{e_i}{\mathbf{D}}{e_i}$ is not a hindrance to the treatment of system-bath coupling for a broad class of thermal baths~\cite{breuerbook}: For dipolar system-bath coupling [$H_\mathrm{SB}=\mathbf{D}\cdot\tilde{\mathbf{B}}$ as in Eq.~\eqref{eq_H_SB}], the static part of the dipole operator in the interaction picture does not contribute to the Liouvillian part of the master equation as long as 
\begin{equation}
\lim_{\omega\rightarrow0}G(\omega)\equiv\lim_{\omega\rightarrow0}\gamma(\omega) (n(\omega)+1)=0,
\end{equation}
which is the case for bosonic baths provided $\gamma(\omega)\propto\omega^\alpha$ with $\alpha>1$. This $\omega=0$ (static) component of the dipole operator, which is absent in the Liouvillian, represents the permanent dipole moment, since (given here for a TLS for brevity)
\begin{equation}
  \mathbf{D}(t)=\mathbf{d}\splus e^{i\omega_0t}+\mathbf{d}^*\sminus e^{-i\omega_0t}+\bkew{e}{\mathbf{D}}{e}\splus\sminus
\end{equation}
with the transition dipole matrix element $\mathbf{d}\coloneq\bkew{e}{\mathbf{D}}{g}$.

\par

Yet, even though Stark-shifted states may realize the aligned-dipoles model (on time scales much shorter than their inverse splitting $1/\Delta$), the predicted power boost is countered by the fact that in an equal superposition of decaying and metastable states only half of the superposed state has dipolar interaction with the bath~\cite{gelbwaser2014power}. Hence, the decay rate is reduced compared to that of the decaying state to $\gamma_\mathrm{eff}=\frac{1}{2}\gamma$. Consequently, the power boost and the reduction in the effective spontaneous emission rate exactly cancel each other. Still, one could argue, that there is a power boost relative to a TLS decaying at rate $\gamma_\mathrm{eff}$.

\par

It has been suggested that one may create an effective $V$-type system with aligned dipoles out of a $\Lambda$-system with orthogonal transitions by applying a strong laser field~\cite{ficek2002quantum}. The resulting $V$-system is non-degenerate, but the energy mismatch can be tuned to be very small, such that one can observe the effects discussed here on long time scales. However, as for the Stark effect discussed above, the decay rate of the dressed state is smaller than that of the bare excited state, precluding a power boost.

\par

In view of these difficulties, the strict alignment requirement should be relaxed in realistic experimental situations. Even though for quasi-aligned dipole transition vectors no dark states exist, strongly subradiant (metastable) states may still arise. The non-thermalization of these ``quasi-dark'' states should therefore be experimentally observable on appropriate time scales.

\par

A promising alternative is to make use of the plethora of vibrational and rotational levels in molecules as discussed in the supplemental material of Ref.~\cite{tscherbul2014long}. 

\subsection{The Dicke system}\label{sec_multiatom_realization}

In view of the conceptual difficulties in finding a realization of parallel dipoles in a degenerate multilevel system that is capable of power boost compared to a TLS, we shall revisit the Dicke system discussed in Sec.~\ref{sec_dicke}. As we have seen, this multipartite setup exhibits similar thermodynamic properties to a degenerate $N$-level system of parallel dipoles and may produce a power boost under suitable initial conditions.

\par

Traditionally, the superradiant Dicke model was thought to be realizable by a dense atomic ensemble confined well within the cubed emission wavelength~\cite{breuerbook,gross1982superradiance}. Yet, in general, such ensembles do not obey the Dicke model, since the dipole-dipole interaction as well as the cooperative decay rates strongly vary, depending on the spatial symmetry and spacings of the atomic ensemble~\cite{lehmberg1970radiation,kurizki1985quantum,kurizki1987theory,mazets2007multiatom,petrosyan2002scalable,scully2009collective}. The remedy is to realize the Dicke model in field-confining structures such as photonic bandgap structures~\cite{kurizki1990two,li2008fabrication}, cavities~\cite{kurizki1996resonant} (see also~\cite{binder2015quantacell}), and waveguides~\cite{shahmoon2011strongly,shahmoon2013nonradiative,shahmoon2013dispersion,shahmoon2014nonlinear}.

\par
\begin{figure}
  \centering
  \includegraphics[width=\columnwidth]{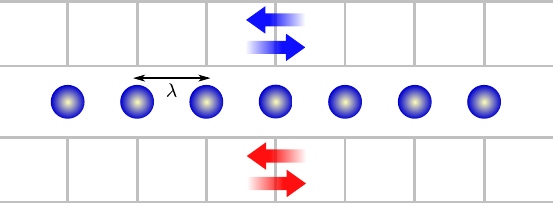}
  \caption{(Color online) A realization of the Dicke model by means of two-level atoms confined at interatomic distance $\lambda$ next to or within one-dimensional waveguides. See text for details.}\label{fig_dicke_realization}
\end{figure}
\par

In particular, consider the realistic setup (Fig.~\ref{fig_dicke_realization})~\cite{mitsch2014quantum} of a periodic $1$d lattice of atoms (each tightly confined by the lattice potential) that are positioned at equal distances $d$ within or next to a photonic waveguide. Alternatively, we may employ a chain of superconducting qubits coupled to a microwave coplanar waveguide~\cite{vanloo2013photon}. Field confinement can give rise to a giantly enhanced density of modes at the lower cutoff frequency of the waveguide~\cite{kleppner1981inhibited,shahmoon2013nonradiative}. Then, atoms whose resonance frequency $\omega_0$ is just above $\omega_\mathrm{cutoff}$ are predominantly coupled to the $1$d axial-mode continuum, whereas transverse-mode (free-space) effects are negligible in comparison, $\gamma_{1\mathrm{d}}\gg\gamma_\mathrm{free}$~\cite{kleppner1981inhibited,shahmoon2013nonradiative}. In such an effective $1$d photonic ``bath'', the cooperative decay rates $\Gamma_{ij}$ and resonant dipole-dipole interactions $\Omega_{ij}$ of atoms $i$ and $j$ are~\cite{vanloo2013photon,lalumiere2013input}
\begin{subequations}\label{eq_dicke_realization_rates_general}
  \begin{align}
    \Gamma_{ij}&=\gamma_{1\mathrm{d}}\cos\left(kd\right)\\
    \Omega_{ij}&=\frac{\gamma_{1\mathrm{d}}}{2}\sin\left(kd\right),
  \end{align}
\end{subequations}
with the wavenumber $k=2\pi/\lambda=\omega_0/c$. The oscillatory behavior of Eqs.~\eqref{eq_dicke_realization_rates_general} stands in stark contrast to the decay with distance of their counterparts in free space~\cite{lehmberg1970radiation} and allows us to choose the interatomic distance $d=\lambda$ such that
\begin{subequations}
  \begin{align}
    \Gamma_{ij}&=\gamma_{1\mathrm{d}}\label{eq_dicke_realization_rates}\\
    \Omega_{ij}&=0.
  \end{align}
\end{subequations}
In this geometry, the atomic lattice realizes the Dicke system~\cite{dicke1954coherence,agarwalbook,gross1982superradiance}, i.e., diagonalization of the $N$-atom Liouvillian operator with the interatomic decay rates~\eqref{eq_dicke_realization_rates} leads to the Liouvillian~\eqref{eq_L_dicke} involving a single decay channel, whose collective steady-state solution is the partial thermal state~\eqref{eq_ss_dicke}. By coupling the atoms to two similar waveguides kept at different temperatures (see Fig.~\ref{fig_dicke_realization}), we may operate the setup as the heat machine depicted in Fig.~\ref{fig_system}a.

\par 

The high-temperature regime, where the Dicke model may outperform the fully thermalized model corresponds to temperatures $\Teff\gtrsim \hbar\omega_0/(5\kB)$ (see Fig.~\ref{fig_currents_dicke}). For optical transitions ($\omega_0\sim10^{15}$\,Hz) these are temperatures of the order of $10^3$\,K. By contrast, for microwave transitions ($\omega_0\sim10^9$\,Hz) the high-temperature regime is already realized at $\Teff\gtrsim10^{-3}$\,K.

\section{Conclusions}\label{sec_conclusions_and_outlook}

The present analysis sheds new light on multilevel or multipartite heat machines and their degeneracy as a thermodynamic resource that can boost the heat currents and the power output of heat engines. By contrast, the efficiency of such machines was shown to be unaltered compared to a TLS-based machine, which adheres to the Carnot bound.

\par

We can summarize our findings regarding the r\^ole of initial and steady-state coherences in the power enhancement of degenerate steady-state multilevel quantum heat engines as follows,
\begin{enumerate}

\item \emph{Steady-state coherences are a consequence of thermalization if at least two transition dipoles are parallel.} Such coherences arise in the bare energy basis but not in the rotated basis containing dark states.

\item \emph{Neither initial nor steady-state coherences do necessarily imply power enhancement compared to a single TLS and, vice versa, power enhancement does not always entail initial or steady-state coherences.} The relevant factor for power enhancement is the thermalization capability (Sec.~\ref{subsec_parallel}) of the initial state, not the amount of steady-state coherence. As an extreme example, even if all three dipoles of a four-level system are perpendicular to each other (the largest orthogonal system possible in three-dimensional space) the power is enhanced---without coherence between the excited states at any time and regardless of the initial condition. The decisive ingredient is the \emph{common ground state} of the dipole transitions. It is generally true that aligned dipole configurations yield higher power boost than non-aligned ones in the high-temperature limit---if the initial state is properly chosen. Even then, we may not attribute this power boost (compared to non-aligned configurations) to steady-state coherences but rather to the presence of dark states, which reduce the number of available levels amenable to thermalization. The key factor for enhancement is the \emph{ground-state population} of this reduced $\Neff$-level system. The fewer levels can thermalize, the more is the ground state thermally populated. Yet, any population in the ground state necessarily \emph{decreases} the maximally permitted coherences within the excited-states manifold. 

\item \emph{Power reduction implies the existence of steady-state coherences.} Power reduction is associated with dark states being initially populated. The largest possible steady-state coherences correspond to fully excited (population-inverted) states, which can only occur if the initial state is dark.

\end{enumerate}

\par

Overall, we found that an $N$-level heat machine cannot yield higher power than $N-1$ independent two-level heat machines. Whilst at low-temperatures both machines perform equally well, the power enhancement in the $N$-level machine decreases at higher temperatures, unless all dipoles are parallel. Taking advantage of the full cooperativity of the excited states (occurring only for parallel transition dipoles), however, requires carefully chosen (non-dark) initial conditions (Sec.~\ref{sec_ss}) and its realization encounters conceptual difficulties (Sec.~\ref{subsec_realizations_parallel}).

\par

Similar conclusions were found to apply in $N$-atom ensembles that, under specially-chosen conditions, obey the Dicke model (Sec.~\ref{sec_multiatom_realization}). Remarkably, the $N$-atom Dicke model yields at best doubly-enhanced power compared to $N$ independent atoms (and only at high temperatures) (Sec.~\ref{sec_dicke}). Nevertheless, doubling the power output of a heat machine by cooperative effects will be an important achievement.

\par

To conclude, our findings elucidate the possible impact of steady-state coherence or entanglement in the working medium on the performance of heat machines. The fact that the heat baths are assumed to be thermal allows us to uphold the Carnot bound. The possibility of exceeding the Carnot bound by virtue of harnessing \emph{non-thermal} baths in the heat machine, including baths consisting of coherently-superposed~\cite{scully2003extracting,abah2014efficiency,turkpence2015quantum} or superradiant~\cite{hardal2015superradiant} atomic systems, as well as squeezed baths~\cite{rossnagel2014nanoscale} is outside the scope of the present discussion, and so are effects related to piston quantization~\cite{quan2012validity,boukobza2013breaking,gelbwaser2013work,gelbwaser2014heat,gelbwaser2015work} or to strong system-bath coupling~\cite{nieuwenhuizen2002statistical,gallego2014thermal,gelbwaser2015strongly} that require further clarification. Notwithstanding these open issues, the present results help delineate the limited (but significant) part that quantum coherence or entanglement may play in quantum thermodynamics.

\begin{acknowledgments}
  We would like to thank Laurin Ostermann for helpful discussions. This work has been supported by the BSF, ISF, AERI, MOST, and CONACYT.
\end{acknowledgments}

\appendix

\section{Coefficients and solution of the ODE for the three-level system}\label{app_threelevels}

The coefficient matrix $\mathcal{A}$ and the inhomogeneity $\mathbf{b}$ appearing in the ODE~\eqref{eq_ode} are (generalizing Ref.~\cite{gelbwaser2014power})
\begin{multline}\label{eq_odes_a}
  \mathcal{A}\coloneq \frac{1}{2}G(\omega_0)\times\\
  \times
  \begin{pmatrix}
    -1-\alpha^2 & 0 & \mathfrak{p}\alpha e^{i\varphi}\left(1+2 e^{-\beta\hbar\omega_0}\right) & 0 \\
    0 & -1-\alpha^2 & \mathfrak{p}\alpha e^{-i\varphi}\left(1+2 e^{-\beta\hbar\omega_0}\right) & 0 \\
    2\mathfrak{p}\alpha e^{-i\varphi} & 2\mathfrak{p}\alpha e^{i\varphi} & -2-2 e^{-\beta\hbar\omega_0} (1+\alpha^2) & 2 (\alpha^2-1) \\
    -\mathfrak{p}\alpha e^{-i\varphi} & -\mathfrak{p}\alpha e^{i\varphi} & 2\alpha^2 e^{-\beta\hbar\omega_0} & -2\alpha^2
  \end{pmatrix}
\end{multline}
and
\begin{equation}
  \mathbf{b}\coloneq \frac{1}{2}G(\omega_0)
  \begin{pmatrix}
    -\mathfrak{p}\alpha e^{i\varphi} \\ -\mathfrak{p}\alpha e^{-i\varphi} \\ 2 \\ 0
  \end{pmatrix}
  .
\end{equation}
Here we have defined the shorthand notation $\mathfrak{p}_{12}\equiv\mathfrak{p}_{21}\eqcolon\mathfrak{p}$, $\varphi_{12}\equiv-\varphi_{21}\eqcolon\varphi$ [\cf Eq.~\eqref{eq_def_pij}], and the dipole strengths are $\alpha_1=1$ and $\alpha_2\eqcolon\alpha$ [\cf Eq.~\eqref{eq_def_alpha}].

\par

The coefficient matrix appearing in the ODE~\eqref{eq_ode_y} explicitly reads
\begin{equation}
  \mathcal{B}\coloneq-\frac{1}{2}G(\omega_0)
  \begin{pmatrix}
    a & 0 & c^* & 0 \\
    0 & a & 0 & c \\
    c & 0 & b & 0 \\
    0 & c^* & 0 & b
  \end{pmatrix}
  ,
\end{equation}
where
\begin{subequations}
  \begin{align}
    a&\coloneq 1+\left(1+\alpha\right)^2e^{-\beta\hbar\omega_0}\\
    b&\coloneq \alpha^2+\left(1+\alpha\right)^2e^{-\beta\hbar\omega_0}\\
    c&\coloneq \mathfrak{p}\alpha e^{i\varphi}.
  \end{align}
\end{subequations}

\par

The general steady-state solution of the linear ODE~\eqref{eq_ode} is then found to be
\begin{subequations}\label{eq_ss_threelevels}
  \begin{align}
    \rho_{21}^\mathrm{ss}&=
    \begin{cases}
      \frac{\alpha e^{i\varphi}}{1+\alpha^2}\frac{1-\mathcal{I}\left(2+e^{\beta\hbar\omega_0}\right)}{1+e^{\beta\hbar\omega_0}}&\quad \mathfrak{p}=1\\
      0&\quad \mathfrak{p}\in[0,1)
    \end{cases}\label{eq_ss_threelevels_coh1}\\
    \rho_{12}^\mathrm{ss}&=\left(\rho_{21}^\mathrm{ss}\right)^*\label{eq_ss_threelevels_coh2}\\
    \rho_{00}^\mathrm{ss}&=
    \begin{cases}
      \frac{1}{1+e^{-\beta\hbar\omega_0}}\left(1-\mathcal{I}\right)&\quad \mathfrak{p}=1\\
      \frac{1}{1+2e^{-\beta\hbar\omega_0}}&\quad \mathfrak{p}\in[0,1)
    \end{cases}\\
    \rho_{22}^\mathrm{ss}&=
    \begin{cases}
      \frac{1}{1+\alpha^2}\frac{\alpha^2+\mathcal{I}\left(1-\alpha^2+e^{\beta\hbar\omega_0}\right)}{1+e^{\beta\hbar\omega_0}}&\quad \mathfrak{p}=1\\
      \frac{1}{2+e^{\beta\hbar\omega_0}}&\quad \mathfrak{p}\in[0,1)
    \end{cases}
    ,
  \end{align}
\end{subequations}
where
\begin{equation}\label{eq_I}
\mathcal{I}\coloneq\frac{1}{1+\alpha^2}\left\{\alpha^2\rho_{11}(0)+\rho_{22}(0)-2\alpha\realt\left[e^{i\varphi}\rho_{12}(0)\right]\right\}
\end{equation}
is an integral of motion associated with the overlap of the initial state with a dark state, to be discussed in Sec.~\ref{sec_ss}.

\section{Proof of the steady-state solution~\eqref{eq_ss_nlevels_pneq1} for non-aligned dipoles}\label{app_proof_pneq1}

\begin{proof}
  The Liouvillian~\eqref{eq_L} for a single heat bath at inverse temperature $\beta$ and without modulation ($q=0$) reads
  \begin{multline}
    \mathcal{L}\rho=\frac{1}{2}G(\omega_0)\sum_{j=1}^{N-1}\left[\alpha_j^2\mathcal{D}_{\downarrow}^{jj}+\sum_{\substack{j^\prime\neq j}}\mathfrak{p}_{jj^\prime}e^{i\varphi_{jj^\prime}}\alpha_j\alpha_{j^\prime}\mathcal{D}_{\downarrow}^{jj^\prime}\right]\\
    +\frac{1}{2}G(-\omega_0)\sum_{j=1}^{N-1}\left[\alpha_j^2\mathcal{D}_{\uparrow}^{jj}+\sum_{\substack{j^\prime\neq j}}\mathfrak{p}_{jj^\prime}e^{-i\varphi_{jj^\prime}}\alpha_j\alpha_{j^\prime}\mathcal{D}_{\uparrow}^{jj^\prime}\right]
  \end{multline}
  with the abbreviations $\mathcal{D}_{\downarrow}^{ij}\coloneq\mathcal{D}\left(\sminus^i,\splus^j\right)$ and $\mathcal{D}_{\uparrow}^{ij}\coloneq\mathcal{D}\left(\splus^i,\sminus^j\right)$. We now apply this operator to the diagonal ansatz
  \begin{equation}
    \rho^\mathrm{ss}=\sum_{k=1}^N\rho_{kk}^\mathrm{ss}\proj{k}.
  \end{equation}
  Making use of the KMS relation~\eqref{eq_kms} we find
  \begin{multline}
    \mathcal{L}\rho^\mathrm{ss}=\frac{1}{2}G(\omega_0)\times\\
    \times\sum_{i=1}^{N-1}\alpha_i^2\left(2\rho_{ii}^\mathrm{ss}-2\rho_{00}^\mathrm{ss}e^{-\beta\hbar\omega_0}\right)\left(\proj{0}-\proj{i}\right)+\\
    +\frac{1}{2}G(\omega_0)\times\\
    \times\sum_{\substack{i,j=1\\i\neq j}}^{N-1}\mathfrak{p}_{ij}\alpha_i\alpha_je^{-i\varphi_{ij}}\left(-\rho_{ii}^\mathrm{ss}-\rho_{jj}^\mathrm{ss}+2\rho_{00}^\mathrm{ss}e^{-\beta\hbar\omega_0}\right)\ketbra{i}{j}.
  \end{multline}
  Hence the diagonal thermal state $\rho_{ii}^\mathrm{ss}=e^{-\beta\hbar\omega_0}\rho_{00}^\mathrm{ss}$ [Eq.~\eqref{eq_ss_nlevels_pneq1}] is indeed the steady-state solution of the master equation $\mathcal{L}\rho^\mathrm{ss}=0$ for $N$ levels when the dipoles are not aligned.
\end{proof}

\section{Proof of the steady-state solution~\eqref{eq_ss_nlevels_p1} for aligned dipoles}\label{app_proof_p1}

\begin{proof}
  The master equation 
  \begin{equation}\label{eq_master_nlevel_p1}
    \dot\rho=\left(\mathcal{L}_\mathrm{decay}+\mathcal{L}_\mathrm{absorption}\right)\rho
  \end{equation}
  associated to the system-bath interaction Hamiltonian~\eqref{eq_H_SB_parallel} is the sum of the two Liouvillians
  \begin{subequations}\label{eq_L_nlevel_p1}
    \begin{align}
      \mathcal{L}_\mathrm{decay}\rho&=\frac{1}{2}G(\omega_0)\left(\sum_{j=1}^{N-1}\alpha_j^2\right)\mathcal{D}\left(\sminusbar,\splusbar\right)\\
      \mathcal{L}_\mathrm{absorption}\rho&=\frac{1}{2}G(-\omega_0)\left(\sum_{j=1}^{N-1}\alpha_j^2\right)\mathcal{D}\left(\splusbar,\sminusbar\right),
    \end{align}
  \end{subequations}
  describing decay and absorption processes mediated by the bath between the bright- and the ground states, respectively.
  
  \par

  We first show that the steady-state solution cannot contain coherences between non-dark and dark states because according to equation~\eqref{eq_L_nlevel_p1} the Ehrenfest equations for these coherences read
  \begin{subequations}\label{eq_proof_coherences_dark_nondark_vanish}
    \begin{gather}
      \frac{\dd}{\dd t}\ew{\ketbra{\psid^k}{\psib}}\propto-\ew{\ketbra{\psid^k}{\psib}}\\
      \frac{\dd}{\dd t}\ew{\ketbra{\psid^k}{0}}\propto-e^{-\beta\hbar\omega_0}\ew{\ketbra{\psid^k}{0}},
    \end{gather}
  \end{subequations}
  and hence $\ew{\ketbra{\psid^k}{\psib}}_\mathrm{ss}=0$ and $\ew{\ketbra{\psid^k}{0}}_\mathrm{ss}=0$. Consequently, the steady-state solution can be decomposed into a dark-state portion and a portion in the subspace orthogonal to the dark states,
  \begin{equation}
    \rho^\mathrm{ss}=\rho_\perp^\mathrm{ss}+\rho_\mathrm{d}^\mathrm{ss}
  \end{equation}
  with the normalizations
  \begin{subequations}
    \begin{align}
      \Tr\rho_\perp^\mathrm{ss}&=1-\Pidew\equiv\bkew{\psib}{\rho(0)}{\psib}+\bkew{0}{\rho(0)}{0}\label{eq_nlevel_p1_normalisation}\\
      \Tr\rho_\mathrm{d}^\mathrm{ss}&=\Pidew\equiv\sum_{j=1}^{N-2}\bkew{\psid^j}{\rho(0)}{\psid^j}.
    \end{align}
  \end{subequations}
  The latter normalization condition follows from the fact that the relative weights of non-dark (bright and ground) states and initial dark-state contributions cannot be altered by the Liouvillian dynamics.
  \par
  We then assume a diagonal steady-state density matrix. Inserting such an ansatz into the master equation~\eqref{eq_L_nlevel_p1} and making use of the KMS relation~\eqref{eq_kms} yields, in the basis spanned by the bright and ground states,
  \begin{multline}
    \mathcal{L}\rho^\mathrm{ss}=\frac{1}{2}G(\omega_0)\left(\sum_{j=1}^{N-1}\alpha_j^2\right)\left[2\rho_{\mathrm{bb}}^\mathrm{ss}-2\rho_{00}^\mathrm{ss}e^{-\beta\hbar\omega_0}\right]\times\\\times\Big[\proj{0}-\proj{\psib}\Big].
  \end{multline}
  The right-hand side obviously gives zero for $\rho_{\mathrm{bb}}^\mathrm{ss}=\rho_{00}^\mathrm{ss}e^{-\beta\hbar\omega_0}$. Combining the latter relation and the normalization condition~\eqref{eq_nlevel_p1_normalisation} yields the populations $\rho_\mathrm{bb}^\mathrm{ss}$ and $\rho^\mathrm{ss}_{00}$ from equation~\eqref{eq_ss_nlevels_p1}.
  \par
  The part of the initial density matrix lying within the dark-state space is not altered by the dynamics and hence $\rho_\mathrm{d}^\mathrm{ss}\equiv\rho_\mathrm{d}(0)$. This state, in principle, can be an arbitrary superposition of the dark basis vectors. We aim at a diagonal representation of the full density matrix, hence we diagonalize the dark-state density matrix and find the relative weights $\rho_{kk}^\mathrm{ss}$, which are the corresponding eigenvalues of $\rho_\mathrm{d}(0)$.
\end{proof}

\section{System-bath interaction Hamiltonian for arbitrary dipole geometry}\label{app_H_SB_arbitrary_geometry}

As shown in Sec.~\ref{subsec_parallel}, parallel dipoles effectively correspond to a single transition with an enhanced transition strength between the ground state and a certain bright state. Accordingly, the interaction of the $l$th ``domain'' ($1\leq l\leq p$) with the bath is described by the effective Hamiltonian
\begin{multline}
  H_\mathrm{SB,\text{ }domain}^l=\sqrt{\sum_{j=n_{l-1}+1}^{n_l}\alpha_j^2}|\mathbf{d}_1|\times\\\times\left(\splusbar^l\otimes \mathbf{e}_l\cdot\mathbf{B} + \sminusbar^l\otimes \mathbf{e}_l^*\cdot\mathbf{B}^\dagger\right)
\end{multline}
with the Pauli operators [\cf Eq.~\eqref{eq_pauli_bright_states}]
\begin{subequations}\label{eq_splusbar}
  \begin{align}
    \splusbar^l&\coloneq\ketbra{\psib^{l}}{0}\\
    \sminusbar^l&\coloneq\ketbra{0}{\psib^{l}}
  \end{align}
\end{subequations}
and the unit vector $\mathbf{e}_l\in\mathbb{C}^3$ pointing along the transition-dipole direction of the $l$th domain.
\par
The interaction of the remaining transitions, which are not parallel to any other, with the bath are described by the Hamiltonians ($N_p+1\leq l\leq N-1$)
\begin{equation}
  H_\mathrm{SB}^l=\alpha_l|\mathbf{d}_1|\left(\splus^l\otimes \mathbf{e}_l\cdot\mathbf{B} + \sminus^l\otimes \mathbf{e}_l^*\cdot\mathbf{B}^\dagger\right),
\end{equation}
with the ``bare'' Pauli operators $\splus^l\coloneq\ketbra{l}{0}$ and $\sminus^l\coloneq\ketbra{0}{l}$ defined in Eq.~\eqref{eq_splus}, and $\mathbf{e}_l$ pointing in the direction of the $l$th dipole transition vector.
\par
The interaction Hamiltonian can thus be decomposed as
\begin{equation}\label{eq_SB_nlevel_general}
  H_\mathrm{SB}=\sum_{l=1}^p H_\mathrm{SB,\text{ }domain}^l+\sum_{l=N_p+1}^{N-1}H_\mathrm{SB}^l.
\end{equation}
It \emph{formally} corresponds to a system with
\begin{equation}
  \Neff\coloneq p+N-N_p
\end{equation}
levels that consist of the $p$ bright states, the $N-1-N_p$ excited states with non-aligned transition dipoles, and the ground state $\ket{0}$. The remaining $\sum_{j=1}^{p}(n_j-1)\equiv N-\Neff$ levels are dark and thus decoupled from the dynamics.

\section{Heat currents and power for the multilevel system}\label{app_heat_currents_power}

The heat currents associated to the $i$th bath ($i\in\{\indexc,\indexh\}$) for the degenerate $N$-level system explicitly read
\begin{multline}\label{eq_heat_currents_general}
  J_i=\sum_{q\in\mathbb{Z}}\hbar(\omega_0+q\Omega)P(q)G_i(\omega_0+q\Omega)\times\\\times\frac{e^{-\beta_i\hbar(\omega_0+q\Omega)}-e^{-\betaeff\hbar\omega_0}}{1+(\Neff-1)e^{-\betaeff\hbar\omega_0}}\left(\sum_{j=1}^{N-1}\alpha_j^2\right)\Big[1-\Pidew\Big].
\end{multline}
Having found the heat currents, we may now compute the power, which according to the first law of thermodynamics (energy conservation) reads $\dot{W}=-(\Jh+\Jc)$. Upon inserting the heat currents~\eqref{eq_heat_currents_general} for $i\in\{\indexc,\indexh\}$ into this equation, we find the power
\begin{multline}\label{eq_power_general}
  \dot{W}=\sum_{q\in\mathbb{Z}}\sum_{i\in\{\indexc,\indexh\}}\hbar(\omega_0+q\Omega)P(q)G_i(\omega_0+q\Omega)\times\\\times\frac{e^{-\betaeff\hbar\omega_0}-e^{-\beta_i\hbar(\omega_0+q\Omega)}}{1+(\Neff-1)e^{-\betaeff\hbar\omega_0}}\left(\sum_{j=1}^{N-1}\alpha_j^2\right)\Big[1-\Pidew\Big].
\end{multline}

\section{Cooperative heat currents for the Dicke system}\label{app_heat_currents_dicke}

For the $N$-atom Dicke system the heat currents read ($i\in\{\indexc,\indexh\}$)
\begin{multline}\label{eq_dicke_heat_currents}
  J_i=\sum_{q\in\mathbb{Z}}N\hbar(\omega_0+q\Omega)P(q)G_i(\omega_0+q\Omega)\left[\sum_{j=0}^{N-1} e^{-j\betaeff\hbar\omega_0}\right]\times\\\times\frac{e^{-\beta_i\hbar(\omega_0+q\Omega)}-e^{-\betaeff\hbar\omega_0}}{\sum_{j=0}^N e^{-j\betaeff\hbar\omega_0}}\Big[1-\Pidew\Big].
\end{multline}
We again observe the dependence of the heat flows on the thermalization capability $1-\Pidew$. These heat currents are governed by three inverse temperatures, $\betac$, $\betah$, and $\betaeff$.

\end{document}